\documentclass[aps,pre,superscriptaddress,twocolumn]{revtex4-1}

\usepackage[utf8]{inputenc}

\usepackage{cancel}
\usepackage{color}

\usepackage{amssymb,amsmath,bm}
\usepackage{graphicx}

\usepackage{hyperref}

\usepackage{mathtools}

\newcommand{\eq}{\text{eq}}
\newcommand{\ini}{\text{i}}
\newcommand{\fin}{\text{f}}
\newcommand{\calW}{W}
\newcommand{\irr}{\text{irr}}

\newcommand{\calH}{\mathcal{H}}
\newcommand{\calK}{\mathcal{K}}

\newcommand{\unb}{\text{u}}
\newcommand{\bou}{\text{b}}
\newcommand{\lin}{\text{lin}}

\newcommand{\pder}[2]{\frac{\partial{#1}}{\partial{#2}}}

\newcommand{\ym}{ y_{\text{m}}}

\def\dbar{{\mathchar'26\mkern-12mu d}}

\begin{document}


\title{Optimal work in a harmonic trap with bounded stiffness}



\author{Carlos A. Plata}
\email[]{cplata1@us.es}
\affiliation{Física Teórica, Universidad de Sevilla, Apartado de
  Correos 1065, E-41080 Sevilla, Spain}
\affiliation{Dipartimento di Fisica e Astronomia “Galileo Galilei”, Istituto Nazionale di Fisica Nucleare,
Università di Padova, Via Marzolo 8, 35131 Padova, Italy}

\author{David Gu\'ery-Odelin}
\email[]{dgo@irsamc.ups-tlse.fr}
\affiliation{Laboratoire de Collisions Agr\'egats R\'eactivit\'e, CNRS, UMR 5589, IRSAMC, France}

\author{E. Trizac}
\email[]{trizac@lptms.u-psud.fr}
\affiliation{LPTMS, UMR 8626, CNRS, Univ. Paris-Sud, Université Paris-Saclay, 91405 Orsay, France}

\author{A. Prados}
\email[]{prados@us.es}
\affiliation{Física Teórica, Universidad de Sevilla, Apartado de
  Correos 1065, E-41080 Sevilla, Spain}


\date{\today}

\begin{abstract}
  We apply Pontryagin's principle to drive rapidly a trapped
  overdamped Brownian particle in contact with a thermal bath between
  two equilibrium states corresponding to different trap stiffness
  $\kappa$. We work out the optimal time dependence $\kappa(t)$ by
  minimising the work performed on the particle under the
  non-holonomic constraint $0\leq\kappa\leq\kappa_{\max}$, an
  experimentally relevant situation.  Several important differences
  arise, as compared with the case of unbounded stiffness that has
  been analysed in the literature. First, two arbitrary equilibrium
  states may not always be connected. Second, depending on the
  operating time $t_{\fin}$ and the desired compression ratio
  $\kappa_{\fin}/\kappa_{\ini}$, different types of solutions
  emerge. Finally, the differences in the minimum value of the work
  brought about by the bounds may become quite large, which may have a
  relevant impact on the optimisation of heat engines.
\end{abstract}


\maketitle


\section{Introduction}\label{sec:intro}


One of the key parameters in non-equilibrium transformations 
is the characteristic relaxation time of the system under study. In general,
equilibrium states of a system depend on the values of certain
physical properties $\lambda$ that can be externally controlled, such
as the available volume for a gas or the spring constant of the
harmonic potential that confines a colloidal particle. When one
relevant external parameter is abruptly changed from $\lambda_{\ini}$
to $\lambda_{\fin}$, a system that was at the equilibrium state
corresponding to $\lambda_{i}$ begins to evolve and, as time
increases, approaches the new equilibrium state corresponding to
$\lambda_{\fin}$. The system's equilibration time $t_{\eq}$ can be
loosely defined as the time that the system needs to reach the new
equilibrium configuration, and it is an intrinsic property for each
physical system that depends on the underlying interactions, 
encoded in the transport coefficients, the external parameters
$\lambda$, and the temperature.

Recently, there has been a growing interest in the development of
engineered techniques capable of beating the natural time scale for
relaxation between equilibrium states. Inspired by the so-called
shortcut to adiabaticity processes
\cite{chen_fast_2010,chen_shortcut_2010}, specific procedures that make
it possible to connect equilibrium states using linking times much
shorter than the natural equilibration time have been devised. The term
\textit{Engineered Swift Equilibration} (ESE) has been coined to
describe these kind of procedures. The general idea of an ESE process
is to design a tailor-made time dependent protocol $\lambda(t)$ for
the externally controlled parameter, such that the system is driven
from the equilibrium state corresponding to $\lambda_{\ini}$ to the
equilibrium state corresponding to $\lambda_{\fin}$ in a finite time
$t_{\fin}$, ideally much shorter than the equilibration time
$t_{\eq}$. Such protocols have been established for an isolated dilute gas confined in a 3D isotropic harmonic 
trap \cite{guery-odelin_nonequilibrium_2014}, and for nano systems in contact with a thermostat both in the over or underdamped 
regime \cite{martinez_engineered_2016,le_cunuder_fast_2016}.


Here, we focus on a colloidal particle confined by a harmonic trap of
stiffness $\kappa$
\cite{martinez_engineered_2016,le_cunuder_fast_2016,rondin_kramers_2017,chupeau_engineered_2018,chupeau_thermal_2018},
the relevant physical quantity is the variance $\langle x^{2}\rangle$
of its position. Initially, the stiffness of the trap is
$\kappa_{\ini}$ and the particle is equilibrated at the temperature
$T$ of the fluid in which it is immersed,
$\langle x^{2}\rangle_{\ini}=k_{B}T/\kappa_{\ini}$, $k_{B}$ being
Boltzmann's constant. Throughout this work, we consider processes in
which the temperature of the bath is kept constant, at difference with
the approach in \cite{chupeau_thermal_2018}. In a STEP process, the
stiffness of the trap is suddenly changed to a different value
$\kappa_{\fin}$ at $t=0^{+}$, and the relaxation of the colloidal
particle to the new equilibrium state is tracked. Basically, its
variance $\langle x^{2}\rangle$ relaxes exponentially to its new
equilibrium value $\langle x^{2}\rangle_{\fin}=k_{B}T/\kappa_{\fin}$
after a characteristic time $t_{\eq}\simeq 3k_{B}T/(\kappa D)$, where
$D$ is the diffusion coefficient.  Alternatively, the system can be
compressed/decompressed isothermally by introducing a suitable time
protocol $\kappa(t)$ for the stiffness that drives the system from the
initial equilibrium state with $\kappa_{\ini}$ to the final
equilibrium state with $\kappa_{\fin}$ in a finite time
$t_{\fin}$. The ESE procedure consists of choosing in a smart way the
stiffness protocol $\kappa(t)$, so that $t_{\fin}\ll t_{\eq}$, thus
beating the system's natural rate of equilibration. For example, the
protocol employed in Ref.~\cite{martinez_engineered_2016} beats the
natural relaxation time by two orders of magnitude,
$t_{\fin}/t_{\eq}\simeq 0.01$.



Once it has been shown that ESE processes are indeed possible, an
optimisation problem arises. There is a wide class of functions
$\lambda(t)$ that connect the initial and final equilibrium states in
a given time $t_{\fin}$. For each of the possible functions
$\lambda(t)$, one can calculate the work performed in the process
$W=\int_{0}^{t_{\fin}}\pder{H}{\lambda}\dot{\lambda}\, dt$, where $H$ is the
Hamiltonian of the system~\cite{sekimoto_stochastic_2010};
mathematically, $W$ is a functional of $\lambda$. Hence the question,
for a given connection time
$t_{\fin}$: what is the \textit{optimal} time evolution $\lambda^{*}(t)$
that minimises (on average) the work $W$?

For the colloidal particle in a harmonic trap, the optimal time
evolution for the stiffness $\kappa^{*}(t)$ has been obtained for
different boundary
conditions~\cite{schmiedl_optimal_2007,schmiedl_efficiency_2008}. The
specific boundary conditions that are adequate for the ESE process
were considered in \cite{schmiedl_efficiency_2008}, in the context of
building a stochastic heat engine. This result has also been rederived
in later works, see for example \cite{aurell_optimal_2011}. The
optimal protocol for the stiffness has finite jumps both at the
initial and final times, $\kappa^{*}(t=0^{+})\neq\kappa_{\ini}$ and
$\kappa^{*}(t_{\fin}^{-})\neq\kappa_{\fin}$. This kind of discontinuity
at the endpoints of the time interval is usual in stochastic
thermodynamics and stem from the ``Lagrangian'' of the considered
variational problem being linear in the ``velocities''
\cite{band_finite_1982}, which is sometimes known as the Miele problem
\cite{tolle_optimization_2012}. This discontinuities can be
regularised by introducing an additional small term in the Lagrangian,
which introduces two boundary layers of finite width at the endpoints
of the time interval that eliminate the finite jumps
\cite{aurell_boundary_2012,muratore-ginanneschi_application_2017}.

The main shortcoming of previous protocols, be they optimal or not,
comes about in decompression processes. Any protocol involving a short
enough time $t_{\fin}$ entails that the stiffness has to be
transiently negative inside a certain time window
\cite{chupeau_engineered_2018,chupeau_thermal_2018}, similarly to the
situation found in other
systems~\cite{torrontegui_chapter_2013,guery-odelin_nonequilibrium_2014}.
The arising of negative values for the stiffness is challenging from
an experimental point of view, since the potential should change from
confining to repulsive. In the usual experimental setups, the
stiffness $\kappa$ of the harmonic potential is always positive and,
in addition, has a certain upper bound $\kappa_{\max}$ depending on
the technique employed---mainly atomic force microscopy (AFM) or laser
optical tweezers (LOT)
\cite{ritort_single-molecule_2006,wen_force_2007,manosas_force_2007,hoffmann_single_2012,marszalek_stretching_2012,rondin_kramers_2017,ciliberto_experiments_2017}---to
implement the harmonic trap. The existence of this upper limit is
related to the validity of the harmonic approximation. The intrinsic
limit of ESE protocols is dictated by the accuracy of the mathematical
model that describes the physical system.

In light of the above remarks, it is relevant to investigate the
optimisation problem of the work described above when the stiffness of
the trap is restricted to a certain interval,
$\kappa\in[0,\kappa_{\max}]$. The existence of an upper bound also
changes the problem, since very high compression ratios
$\kappa(t)/\kappa_{\ini}$ have to be applied to accelerate the
equilibration in the compression case. For example, in
Ref.~\cite{martinez_engineered_2016}, transient compression ratios of
the order of $40$ were applied in order to speed up the
equilibration of the particle, even when $\kappa_{\fin}$ only doubled
$\kappa_{\ini}$.

These drawbacks are important for the optimisation of irreversible
heat engines, a field of research that has become quite active in the
last few years
\cite{esposito_efficiency_2010,rosnagel_nanoscale_2014,martinez_brownian_2016,martinez_colloidal_2017,taye_irreversible_2017,apertet_true_2017}.
In fact, Brownian particles trapped by optical tweezers have been
recently employed to build stochastic heat engines, both theoretically
and experimentally
\cite{schmiedl_efficiency_2008,blickle_realization_2012,martinez_brownian_2016},
for a review see \cite{ciliberto_experiments_2017}. In these studies,
the stiffness of the trap is changed as a function of time by tuning
the laser power, and decreasing (resp.~increasing) the stiffness is
equivalent to decompressing (resp.~compressing) the system. Cyclic
engines are thus built by connecting isothermal
compression/decompression branches with either isochoric
\cite{blickle_realization_2012} or isoentropic branches
\cite{schmiedl_efficiency_2008,blickle_realization_2012}. In the
decompression (resp.~compression) branch the corresponding work
$W_{d}$ (resp.~$W_{c}$) is negative (resp.~positive), and the total
work $W=W_c+W_d$ must be negative to build a heat engine.

In this work, we focus on the analysis of isothermal
compression/decompression processes, i.e. the isothermal branches of
the heat engines described in the previous paragraph. Note that the
optimisation of the work considered here is relevant in the context of
heat engines, since the extracted work $-W$ has to be a maximum, i.e
$W$ must be minimum \cite{schmiedl_efficiency_2008}.  In addition, the
stiffness is restricted in experiments to a certain interval as
explained above, and thus the externally controlled function
$\kappa(t)$ obeys the non-holonomic constraint
$0\leq \kappa\leq \kappa_{\max}$. Therefore, the currently available
``unconstrained'' results
\cite{schmiedl_efficiency_2008,aurell_optimal_2011} are not useful for
short enough times $t_{\fin}$, because the optimal $\kappa(t)$ becomes
negative (resp.~larger than $\kappa_{\max}$) in decompression
(resp.~compression) processes.


The time evolution of the colloidal particle is governed by a
first-order differential equation,
\begin{equation}\label{eq:x2-ode}
  \frac{d\langle x^{2}(t)\rangle}{dt}=\varphi(\langle x^{2}(t)\rangle,\kappa(t)),
\end{equation}
where $\varphi$ is a smooth function of both $\langle x^{2}\rangle$
and $\kappa$, see for
example~\cite{schmiedl_efficiency_2008,martinez_engineered_2016}. Then,
$\kappa(t)$ is a control function, in the sense used in control
theory. The mean work in a finite time isothermal process can be
written as
\begin{equation}
  W=\frac{1}{2}\int_{0}^{t_{\fin}}dt \,\langle x^{2}\rangle\,
  \dot{\kappa}(t)=-\frac{1}{2}\int_{0}^{t_{\fin}}dt\,\kappa(t)
 \varphi(\langle x^{2}(t)\rangle,\kappa(t)),
\end{equation}
where we have made use of the relation
$\kappa_{\ini}\langle x^{2}\rangle_{\ini}=\kappa_{\fin}\langle
x^{2}\rangle_{\fin}=k_{B}T$. By defining
\begin{equation}
  L(\langle x^{2}\rangle,\kappa)=-\frac{\kappa}{2} \varphi(\langle x^{2}\rangle,\kappa),
\end{equation}
we can write
\begin{equation}
  W=\int_{0}^{t_{\fin}}dt\, L(\langle x^{2}(t)\rangle,\kappa(t)).
\end{equation}
We then have a well-posed problem in control theory
\cite{pontryagin_mathematical_1987,liberzon_calculus_2012}. We seek
the minimum of $W$, taking into account that the evolution of
$\langle x^{2}\rangle$ is controlled by $\kappa$, as described by
\eqref{eq:x2-ode}, where $\kappa(t)$ satisfies the non-holonomic
constraint
\begin{equation}\label{eq:non-holonomic}
  0\leq \kappa(t)\leq \kappa_{\max}.
\end{equation}
This kind of optimisation problem cannot be tackled with the usual
tools of variational calculus, i.e. the Euler-Lagrange equations; they
must be addressed by applying more sophisticated tools from control
theory, such as Pontryagin's maximum
principle~\cite{pontryagin_mathematical_1987,liberzon_calculus_2012}.


The plan of the paper is as follows. Section~\ref{sec:problem} is
devoted to the statement of the minimisation of the work as a control
problem. Therein, we explain how Pontryagin's principle can be applied
to this particular situation. In Sec.~\ref{sec:unbounded}, we address
the minimisation problem when the stiffness is not bounded and can
thus have any value, including negative ones. Next, we look into the
minimisation problem with bounds in Sec.~\ref{sec:bounded}, first for
the decompression case in \ref{sec:decomp} and afterwards for the
compression case in \ref{sec:comp}. Section \ref{sec:phase-diag}
discusses the different phases that appear in the minimisation problem
and a detailed comparison between the values of the optimal work for
the unbounded and the bounded cases is carried out. The main
conclusions are presented in Sec.~\ref{sec:conclusions}. Finally, the
appendices deal with some technicalities that are omitted in the main
text.

\section{The control problem}
\label{sec:problem}

\subsection{Statement}
We consider a colloidal particle immersed in a fluid at temperature
$T$. The particle is in a harmonic trap of stiffness $\kappa(t)$, the time
dependence of which is externally controlled, and we are interested in
time scales such that the overdamped limit holds. Thus, the dynamics
of the particle position $x$ is governed by the Langevin equation
\begin{equation}\label{eq:Langevin}
  \gamma \frac{dx(t)}{dt}=-\kappa(t) x(t) +\xi(t),
\end{equation}
where $\gamma$ is the friction coefficient and $\xi(t)$ is a Gaussian
white noise force,
\begin{equation}\label{eq:GWN}
  \langle \xi(t)\rangle=0, \qquad \langle \xi(t)\xi(t')\rangle=2D\delta(t-t'),
\end{equation}
in which $D$ is the diffusion coefficient that is connected to
$\gamma$ by the fluctuation-dissipation relation $D=k_{B}T/\gamma$.
Implicitly, our modelling assumes that the relaxation of the
surrounding fluid to equilibrium can be regarded as instantaneous on
the time scale over which the stiffness varies.

The Fokker-Planck equation associated to the Langevin equation
\eqref{eq:Langevin} is linear. Therefore, in the class of ESE
processes described in the introduction, the probability
distribution function $\rho(x,t)$ is Gaussian for all times, since it
is so initially, and we can characterise the stochastic process
completely by its variance $\langle x^{2}(t)\rangle$.  To do the
calculations, it is convenient to introduce dimensionless variables
\begin{equation}
  \hat{\kappa}=\frac{\kappa}{\kappa_{\ini}}, \quad
  \hat{t}=\frac{\kappa_{\ini}}{\gamma} t, \quad
  \hat{y}_{1}=\sqrt{\frac{\langle x^{2}\rangle}{\langle
      x^{2}\rangle_{\ini}}},
\end{equation}
where the initial value of the variance is $\langle
x^{2}\rangle_{\ini}=k_{B}T/\kappa_{\ini}$. Therefore, $\hat{y}_{1}(t)$
is the non-dimensional standard deviation. In order not to clutter our
formulas, we omit the hats in the dimensionless variables
henceforth.

The  time evolution of the standard deviation $y_{1}$ is
governed by the first-order differential equation
\begin{subequations}\label{eq:y-evol-with-f1}
  \begin{equation}\label{eq:y-evol}
    \frac{dy_{1}(t)}{dt}=f_{1}(y_{1}(t),\kappa(t)),
  \end{equation}
  with
  \begin{equation}\label{eq:f1-def}
    f_{1}(y_{1},\kappa)\equiv \frac{1}{y_{1}}-\kappa y_{1},
  \end{equation}  
\end{subequations}
for each given time-dependent stiffness $\kappa(t)$.

The mean work performed on the system is defined at the average level as
$\,\dbar W =\frac{1}{2}\langle x^{2}\rangle\,
d\kappa$~\cite{sekimoto_stochastic_2010}, which is positive when
energy is transferred from the environment to the particle and
negative otherwise. The unit of energy is $k_{B}T$, then the
dimensionless work for a finite transformation from $t=0$ to
$t=t_{\fin}$ is, after using integration by
parts~\cite{schmiedl_optimal_2007}
\begin{equation}\label{eq:work-deltaF-plus-irrev}
  \calW =\frac{1}{2}\ln \kappa_{\fin}
  +\int_{0}^{t_{\fin}} dt \, \left[f_{1}(y_{1}(t),\kappa(t))\right]^{2},
\end{equation}
where we have made use of the \textit{boundary conditions} for our
ESE problem
\begin{subequations}\label{eq:bc}
  \begin{equation}\label{eq:bc-k}
    \kappa(0)=\kappa_{\ini}=1, \quad \kappa(t_{\fin})=\kappa_{\fin},
  \end{equation}
  \begin{equation}\label{eq:bc-y}
  y_{1}(0)\equiv y_{1,\ini}= 1, \quad
  y_{1}(t_{\fin})\equiv y_{1,\fin}=\frac{1}{\sqrt{\kappa_{\fin}}}.
  \end{equation}
\end{subequations}
The first term on the rhs of \eqref{eq:work-deltaF-plus-irrev} is the
free energy difference between the initial and final states. Then, the
second term on the rhs, which is non-negative, is the irreversible
work and vanishes only in the quasi-static limit, when
$t_{\fin}\to\infty$~\cite{schmiedl_optimal_2007,schmiedl_efficiency_2008}.

Here, we are interested in minimising $\calW$ (i.e. maximising the
``extracted'' work $-\calW$) for a fixed time interval $t_{\fin}$,
starting from the equilibrium state corresponding to $\kappa_{\ini}$,
equal to unity in dimensionless variables,  and
ending up in the equilibrium state corresponding to
$\kappa_{\fin}$. Therefore, we have to minimise the irreversible work as
given by the functional
\begin{subequations}\label{eq:functional-def}
  \begin{equation}
    \calW_{\irr}[y_{1},\kappa]=\int_{0}^{t_{\fin}} dt\, f_{0}(y_{1}(t),\kappa(t)),
  \end{equation}
  \begin{equation}
  f_{0}(y_{1},\kappa)\equiv \left[f_{1}(y_{1},\kappa)\right]^{2},
  \end{equation}
\end{subequations}
where the stiffness of the trap $\kappa(t)$ is an externally controlled
function and the time evolution of $y_{1}(t)$ is linked thereto by
\eqref{eq:y-evol}. For the ESE processes, we are especially interested
in the regime
\begin{equation}\label{eq:teq}
t_{\fin}<t_{\eq}, \quad t_{\eq}\simeq\frac{3}{\kappa_{\fin}},
\end{equation}
where $t_{\eq}$ is the equilibration time when the system relaxes
to equilibrium with time-independent stiffness,
$\kappa(t)=\kappa_{\fin}$ for all
times~\cite{martinez_engineered_2016}.

Let us be more specific. For each time-dependent control function
$\kappa(t)$, we obtain a certain time evolution for $y_{1}(t)$ by
integrating \eqref{eq:y-evol}, and therefore a certain value for our
functional $\calW_{\irr}$. What we are interested in is finding out
whether there is an optimal control function $\kappa^{*}(t)$, for
which the corresponding time evolution of the standard deviation is
$y_{1}^{*}(t)$, such that
$\calW_{\irr}[y_{1},\kappa]\geq \calW_{\irr}^{*}\equiv
\calW_{\irr}[y_{1}^{*},\kappa^{*}]$ within a certain class $\calK$ of
\textit{admissible} control functions. From a physical point of view,
it is reasonable to admit functions $\kappa(t)$ with finite
instantaneous jumps at certain times $t\in[0,t_{\fin}]$; therefore we
assume that $\kappa(t)$ is piecewise continuous in
$[0,t_{\fin}]$. Note that this entails that $y_{1}(t)$ must be
continuous in $[0,t_{\fin}]$ since Eq.~\eqref{eq:y-evol} implies that
$\dot{y}_{1}$ has at most finite jump discontinuities.

The boundary conditions for our minimisation problem stem from the ESE
process we are interested in, and are given by \eqref{eq:bc-k}.  At
this point, we have a well-posed \textit{optimal control problem}
\cite{pontryagin_mathematical_1987,gelfand_calculus_2000,liberzon_calculus_2012}. We
want to minimise the functional \eqref{eq:functional-def}, in which
the time evolution of $y_{1}(t)$ is controlled by the imposed program
$\kappa(t)$ by means of the evolution equation \eqref{eq:y-evol}, with
the boundary conditions for $y_{1}$ given by \eqref{eq:bc-y}. This
minimisation is done over the class of admissible controls: piecewise
continuous functions $\kappa(t)$ that verify the prescribed boundary
conditions for $\kappa$, as given by \eqref{eq:bc-k}. In addition, we
may have more restrictions on $\kappa$, which we summarise here by
saying that the possible values of the control $\kappa(t)\in{U}$. The
so-called control set $U$ is a certain subset (interval) of the real
numbers, ${U}\subseteq\mathbb{R}$. Although our notation does not make
it explicit, the control set ${U}$ can vary in time, see for example
section $3.3$ of~\cite{liberzon_calculus_2012}.

\subsection{Pontryagin's procedure}

The solution to this control problem is obtained by applying
Pontryagin's principle, see section $1.8$
of~\cite{pontryagin_mathematical_1987} or section $4.3.1$
of~\cite{liberzon_calculus_2012} for its general formulation.  Below,
we explain how Pontryagin's maximum principle is applied to our
particular physical situation.

First, we define a variable $y_{0}$ such that $y_{0}(0)=0$ and
\begin{equation}\label{eq:y0}
  \frac{dy_{0}}{dt}=f_{0}(y_{1}(t),\kappa(t))=\left[
    f_{1}(y_{1}(t),\kappa(t))\right]^{2}.
\end{equation}
It is clear that, for each choice of the control function $\kappa(t)$,
$y_{0}(t_{\fin})$ equals the value of the functional
$\calW_{\irr}[y,\kappa]$. Next, we introduce variables $\psi_{i}$
conjugate to each $y_{i}$, $i=0,1$, and define a function
\begin{eqnarray}\label{eq:Pi-def}
  \Pi(y_{1},\psi_{0},\psi_{1},\kappa)&=&\psi_{0}f_{0}(y_{1},\kappa)+
                                         \psi_{1}f_{1}(y_{1},\kappa) \nonumber \\
                                     &=& \psi_{0}\left[f_{1}(y_{1},\kappa)\right]^{2}+
                          \psi_{1}f_{1}(y_{1},\kappa).                
\end{eqnarray}
Note that, by construction, $\Pi$ does not depend on
$y_{0}$. For fixed $(y_{1},\psi_{0},\psi_{1})$, the function $\Pi$
becomes a function of $\kappa$, which belongs to the control set, $\kappa\in{U}$. We denote the supremum of this
function by  $\calH$,
\begin{eqnarray}
  \calH(y_{1},\psi_{0},\psi_{1})& = &                      \sup_{\kappa\in{U}}\Pi(y_{1},\psi_{0},\psi_{1},\kappa).            
        \label{eq:H-def}
\end{eqnarray}

In conjunction with \eqref{eq:Pi-def}, the following system of
equations hold for the variables $(y_{0},y_{1},\psi_{0},\psi_{1})$
\begin{equation}\label{eq:canonical-gen}
  \frac{dy_{i}}{dt}=\pder{\Pi}{\psi_{i}}, \qquad
  \frac{d\psi_{i}}{dt}=-\pder{\Pi}{y_{i}}, \quad i=0,1,
\end{equation}
i.e. we recover \eqref{eq:y0} and \eqref{eq:y-evol} for the evolution
of $(y_{0},y_{1})$ and obtain the evolution equations for the
conjugate variables $(\psi_{0},\psi_{1})$
\begin{subequations}\label{eq:canonical-partic}
  \begin{align}\label{eq:canonical-partic-psi}
    &\frac{d\psi_{0}}{dt}=0  \quad \Rightarrow \quad  \psi_{0}(t)=\psi_{0} \text{ (constant),} \\
    &\frac{d\psi_{1}}{dt}=-\psi_{0}\pder{f_{0}}{y_{1}}-
      \psi_{1}\pder{f_{1}}{y_{1}}=-\pder{f_{1}}{y_{1}}\left(2 \psi_{0}f_{1}+\psi_{1}\right).
  \end{align}
\end{subequations}
For any control function $\kappa(t)$ linking $y_{1,\ini}$ and $y_{1,\fin}$
in a time $t_{\fin}$, we have a solution $y_{1}(t)$ of
\eqref{eq:y-evol}. Inserting both $\kappa(t)$ and the associated
$y_{1}(t)$ into \eqref{eq:canonical-partic}, we also obtain the
solutions for the conjugate variables $(\psi_{0},\psi_{1}(t))$
associated to the considered control. This construction defines the conjugate variables,
and consequently the function $\Pi$.

Pontryagin's extremum principle states a necessary condition for
having an optimal control $\kappa^{*}(t)$ that minimises the
functional $\calW$, within the considered class of admissible
controls. Let $\kappa^{*}(t)$ be an admissible control and
$y_{1}^{*}(t)$ the associated solution of \eqref{eq:y-evol}.  In order
that $\kappa^{*}(t)$ yield a solution of the minimisation problem,
there must exist a solution of \eqref{eq:canonical-partic}
$(\psi_{0}^{*},\psi_{1}^{*}(t))\neq (0,0)$ for all
$t\in [0,t_{\fin}]$ such that
\begin{enumerate}
\item for all $t\in[0,t_{\fin}]$, it is at the point
  $\kappa=\kappa^{*}(t)$ that the function
  $\Pi(y_{1}^{*}(t),\psi_{0}^{*}(t),\psi_{1}^{*}(t),\kappa)$ attains
  its maximum, i.e.
  $$ \Pi(y_{1}^{*}(t),\psi_{0}^{*}(t),\psi_{1}^{*}(t),\kappa^{*}(t))=
  \calH(y_{1}^{*}(t),\psi_{0}^{*}(t),\psi_{1}^{*}(t)).
  $$
\item
  The constant $\psi_{0}^{*}\leq 0$. 
\end{enumerate}
The latter condition assures that $\Pi$ has a maximum at
$\kappa^{*}$~\footnote{The main point is that the sign of all the
  momenta $\psi_k$ and thus the sign of $\Pi$ can be reversed, which
  gives a ``mirrored'' solution of the canonical equations. Over this
  ``mirrored'' solution, with $(-\psi_0)>0$, the corresponding
  $(-\Pi)$ would reach an infimum at $(-\calH)$, instead of a
  supremum. It is to fix this ambiguity in Pontryagin’s procedure
  and formulate a maximum principle that the choice $\psi_0<0$ is
  made~\cite{liberzon_calculus_2012}.}. The idea behind Pontryagin's
principle is to rewrite the functional to be extremalised as
$\int_0^{t_\fin} dt\, \psi_0 f_0 = \int_0^{t_\fin} dt\, (\Pi- \psi_1
f_1)$. Taking advantage of the Hamiltonian structure behind
(\ref{eq:canonical-gen}) yields the formalism in question.

From the optimal control, one deduces the corresponding $y_{0}^{*}(t)$
and the minimum irreversible work is
\begin{equation}
  \calW_{\irr}^{\min}=y_{0}^{*}(t_{\fin}).
\end{equation}
Finally, it is straightforward to show that
$\calH(y_{1}^{*}(t),\psi_{0}^{*}(t),\psi_{1}^{*}(t))$ does not depend
on time, i.e. it is a constant of motion.

At this point, the issue is finding the supremum of the function
$\Pi(y_{1},\psi_{0},\psi_{1},\kappa)$ that leads to the optimal
control $\kappa^{*}(t)$. The basic idea is that, for any time $t$, the
value of the optimal control $\kappa$
 can lie either inside $U$ or along its boundary $\partial U$. This is
completely analogous to the situation found when seeking an extremum
of a function of several variables $g(x_{1},x_{2},\ldots,x_{N})$ in a
certain closed subset ${U}\subset\mathbb{R}^{N}$, which may lie
inside ${U}$ or on its boundary $\partial{U}$. To find it, first we look
for the extremum $(x_{1}^{*},x_{2}^{*},\ldots,x_{N}^{*})$ by imposing
$(\partial g/\partial x_{i})^{*}=0$; if this equation does not have a
solution inside ${U}$, the extremum must lie on the boundary
$\partial{U}$. Therefore, to obtain the supremum of $\Pi$, at first
$\kappa^{*}$ is sought by writing
\begin{eqnarray}
  0=\left.\pder{\Pi}{\kappa}\right|_{\tilde{\kappa}}&=&
 \left(\psi_{0}\pder{f_{0}}{\kappa}+
 \psi_{1}\pder{f_{1}}{\kappa}\right)_{\tilde{\kappa}} \nonumber \\
                                &=&\left(\pder{f_{1}}{\kappa}\right)_{\tilde{\kappa}}
 \left(2\psi_{0}f_{1}+\psi_{1}\right)_{\tilde{\kappa}}
  \label{eq:extremum-condition}
\end{eqnarray}
We have introduced the notation $\tilde{\kappa}$ to make it clear that
$\tilde{\kappa}$ may be the ``right'' solution,
i.e. $\tilde{\kappa}=\kappa^{*}$, or not. Being more concrete, there
appear two possibilities:
\begin{enumerate}
\item The specific $\tilde{\kappa}$ found from
  \eqref{eq:extremum-condition} belongs to the class of admissible
  controls for all times $t$, then we have found the solution of the
  minimisation problem, $\kappa^{*}=\tilde{\kappa}$.
\item\label{item:2} $\tilde{\kappa}$ does not belong to the
  class of admissible controls because at a certain time
  $t_{0}<t_{\fin}$ we have that $\tilde{\kappa}(t_{0})$ lies outside
  the control set $U$. Then, the optimal $\kappa^{*}(t)$ comprises in
  general several branches: some branches stem from
  \eqref{eq:extremum-condition} and lie inside $U$ whereas other
  branches lie over its boundary $\partial{U}$.
\end{enumerate}

Now we derive some specific expressions for our system. First, we
write the particular evolution equation for the conjugate variable
$\psi_{1}$,
\begin{equation}\label{eq:psi1-evol-partic}
\frac{d\psi_{1}}{dt}=\left(\frac{1}{y_{1}^{2}}+\kappa\right)
        \left[2\psi_{0}\left(\frac{1}{y_{1}}-\kappa
        y_{1}\right)+\psi_{1}\right],
\end{equation}
where we have taken into account the definition of $f_{1}(y,\kappa)$
in \eqref{eq:f1-def}. Second, we derive the particular equation for
$\tilde{\kappa}$.  Making use of \eqref{eq:extremum-condition} and the
definition of $f_{1}(y,\kappa)$,
\begin{equation}
  0=\left.\pder{\Pi}{\kappa}\right|_{\tilde{\kappa}}=
  -y_{1} \left[2 \psi_{0}
  \left( \frac{1}{y_{1}}-\tilde{\kappa}\,y_{1}\right)+\psi_{1}\right].
  \label{eq:extremum-condition-bis}
\end{equation}
and thus
\begin{equation}\label{eq:tilde-k}
  \tilde{\kappa}=\frac{\psi_{1}}
  {2\psi_{0}\,y_{1}}
  +\frac{1}{y_{1}^{2}}.
\end{equation}
The insertion of \eqref{eq:tilde-k} into the set of differential
equations \eqref{eq:canonical-gen} yields 
\begin{subequations}\label{eq:canonical-unb}
\begin{align}
  & \frac{dy_{0}}{dt}=\left(\frac{\psi_{1}}{2\psi_{0}}\right)^{2},
  & & \frac{dy_{1}}{dt}=-\frac{\psi_{1}}{2\psi_{0}},
  \\
  &\frac{d\psi_{0}}{dt}=0, & &
         \frac{d\psi_{1}}{dt}=0.                                 
\end{align}
\end{subequations}

In the following sections, we analyse in depth two particular cases:
(i) when the stiffness may have any value including negative ones, see
Section \ref{sec:unbounded}, and (ii) when the stiffness is bounded
and lies within a certain interval $[0,\kappa_{\max}]$, see Section
\ref{sec:bounded}. Note that the latter is the relevant problem at the
experimental level, as explained in the introduction.

\section{Unbounded stiffness}\label{sec:unbounded}

First, we consider the simplest situation: we have no other
restrictions on the control function $\kappa(t)$ aside from the
boundary conditions \eqref{eq:bc-k}. Therefore, the class of
admissible control functions $\calK$ comprises all piecewise
continuous functions lying inside the vertical strip
$S_{\unb}\equiv[0,t_{\fin}]\times(-\infty,+\infty)$ in the
$(t,\kappa)$ plane that go from the point $(0,\kappa_{\ini})$ to
$(t_{\fin},\kappa_{\fin})$.

Our starting point is the system of equations
\eqref{eq:canonical-unb}. We add subscripts $\unb$ to all the
variables to mark that we are studying the unbounded case. Both
$\psi_{0,\unb}$ and $\psi_{1,\unb}$ are constants of motion and thus
$y_{1,\unb}$ has a linear shape. The boundary conditions for $y_{1}$,
as given by \eqref{eq:bc-y}, entail that the constant slope equals
$(y_{1,\fin}-y_{1,\ini})/t_{\fin}$, i.e.
\begin{equation}
  \frac{\psi_{1,\unb}}{2\psi_{0,\unb}}=-
  \frac{y_{1,\fin}-1}{t_{\fin}},
\end{equation}
and 
\begin{equation}\label{eq:y1-linear}
  y_{1,\unb}(t)=1+\frac{y_{1,\fin}-1}{t_{\fin}}t.
\end{equation}
In addition,
\begin{equation}
  y_{0,\unb}(t)=\frac{(y_{1,\fin}-1)^{2}}{t_{\fin}^{2}}t.
\end{equation}

Within the theoretical framework of Pontryagin's maximum principle,
the above solution is valid as long as $\tilde{\kappa}$
stemming from \eqref{eq:tilde-k},
\begin{equation}\label{eq:k-unb}
  \tilde{\kappa}_{\unb}(t)=\frac{1}
  {[y_{1,\unb}(t)]^{2}}-\frac{y_{1,\fin}-1}
  {t_{\fin}}\frac{1}{y_{1,\unb}(t)}
\end{equation}
belongs to the class of admissible controls. It can be easily shown
that $\tilde{\kappa}_{\unb}(t)\leq 1$ (resp.~$\geq 1$) for decompression
(resp.~compression). Note that, however, $\tilde{\kappa}_{\unb}$ may
become negative (resp.~arbitrarily large) for decompression
(resp.~compression) as $t_{\fin}$ is reduced.

As already stated at the beginning of this
section, the class of admissible controls for the unbounded case
contains all piecewise functions in the closed interval $[0,t_{\fin}]$
that verify the boundary conditions \eqref{eq:bc-k}. Therefore, the
obtained expression $\tilde{\kappa}_{\unb}(t)$ gives the optimal
control $\kappa^{*}(t)$ in the open interval $(0,t_{\fin})$ but not at
the initial and final times. Therein, $\kappa$ is restricted to only
one value, $\kappa_{\ini}$ for $t=0$ and $\kappa_{\fin}$ for
$t=t_{\fin}$, so it is straightforward that the respective maximums of
$\Pi$ are attained at $\kappa^{*}(0)=\kappa_{\ini}$ and
$\kappa^{*}(t_{\fin})=\kappa_{\fin}$ \footnote{This can also be
  understood as having a time dependent control set ${U}$,
  ${U}(0)=\kappa_{\ini}$, ${U}(t)=[0,\kappa_{\max}]$ for
  $t\in(0,t_{\fin})$ and ${U}(t_{\fin})=\kappa_{\fin}$.}. However,
this poses no problem because the controls have been assumed to be
piecewise continuous in our theory. Therefore, the final result for
the optimal control in the unbounded case is
\begin{equation}
  \kappa_{\unb}^{*}(t)=
    \begin{cases}
      \kappa_{\ini}, &  t=0, \\
      \tilde{\kappa}_{\unb}(t) &  0<t<t_{\fin}, \\
      \kappa_{\fin}, &  t=t_{\fin}.
    \end{cases}
\end{equation}

The optimal profiles for the variables are
$y_{0}^{*}(t)=y_{0,\unb}(t)$ and $y_{1}^{*}(t)=y_{1,\unb}(t)$, with
$[0,t_{\fin}]$.  Neither of them is affected by the finite jumps in
$\kappa_{\unb}^{*}(t)$, since they are continuous functions of
time. Then, we have that
\begin{equation}\label{eq:opt-work-unb}
  \calW_{\irr,\unb}^{*}=y_{0,\unb}(t_{\fin})=\frac{(y_{1,\fin}-1)^{2}}
  {t_{\fin}}=\frac{\left(1-\sqrt{\kappa_{\fin}}\right)^{2}}
  {\kappa_{\fin}\,t_{\fin}}.
\end{equation}
The above results for the optimal standard deviation and the minimum
irreversible work have already been obtained
\cite{schmiedl_efficiency_2008,aurell_optimal_2011}. 

We would like to emphasise the important role played by the boundary conditions to
write the relevant variational problem for the physical situation at
hand. In the context of ESE processes, one wants to connect the
equilibrium states corresponding to $\kappa_{\ini}$ and
$\kappa_{\fin}$ in a finite time $t_{\fin}$ and, therefore, the right boundary
conditions are those given by \eqref{eq:bc}.  Indeed, this is an
important issue that affects the result of the variational problem. For example, the boundary conditions considered
in Ref.~\cite{schmiedl_optimal_2007} do not connect equilibrium states
because the system is not equilibrated at the final time,
$\dot{y}(t_{\fin})\neq 0$.  In fact, this shortcoming was corrected in
Ref.~\cite{schmiedl_efficiency_2008}.

As already stated in the introduction, discontinuities of the optimal
stiffness at the initial and final times often appear in stochastic
thermodynamics
\cite{band_finite_1982,schmiedl_optimal_2007,schmiedl_efficiency_2008,aurell_optimal_2011}.
They are usually rationalised in a mathematical way
\cite{band_finite_1982}, referring to the so-called Miele problem in
which the ``Lagrangian'' is linear in the highest derivative
\cite{tolle_optimization_2012}. We put forward an alternative,
physically appealing, picture to understand the emergence of these
discontinuities in Appendix \ref{app:surgery}.


\section{Bounded stiffness}\label{sec:bounded}

In experiments, the stiffness of the harmonic trap cannot have an
arbitrary value. As stated in the introduction, the type of device
employed to design the harmonic potential (AFM, LOT,\ldots) constrains
the stiffness values to a certain interval
\begin{equation}\label{eq:k-inequality}
  0 \leq \kappa \leq \kappa_{\max}.
\end{equation}
For the sake of concreteness and simplicity, we have taken the minimum
stiffness as $0$ throughout this work. A more general situation with
a non-zero $\kappa_{\text{min}}$ can be addressed along similar lines as  
here. However, note that the most important restriction from
a physical point of view is the positiveness of $\kappa$ which, in
addition, leads to simpler calculations. 

We now turn our attention to the problem of minimising the
irreversible work with the non-holonomic constraint
\eqref{eq:k-inequality}. In this case, the class of admissible control
functions $\calK$ comprises all the piecewise continuous functions
lying inside the rectangle
$S_{\bou}\equiv[0,t_{\fin}]\times[0,\kappa_{\max}]$ in the
$(t,\kappa)$ plane that go from the point $(0,\kappa_{\ini})$ to
$(t_{\fin},\kappa_{\fin})$. Evidently, both $\kappa_{\ini}$ and
$\kappa_{\fin}$ must lie in the interval $[0,\kappa_{\max}]$. The
maximum value of the stiffness $\kappa_{\max}$ leads to a minimum
equilibrium value for the standard deviation, namely
\begin{equation}\label{eq:ymin}
  \ym=\frac{1}{\sqrt{\kappa_{\max}}}.
\end{equation}

Pontryagin's maximum principle is specially adequate to analyse
problems with the kind of non-holonomic constraint in
\eqref{eq:k-inequality}. The condition
$(\partial\Pi/\partial \kappa)_{\tilde{\kappa}}=0$ gives results that
are identical to the unbounded case as long as the protocol
$\kappa^{*}_{\unb}(t)$ lies inside the rectangle
$[0,t_{\fin}]\times[0,\kappa_{\max}]$. When the optimal protocol for
the unbounded case $\kappa^{*}_{\unb}(t)$ crosses the boundary of
this rectangle at a certain time $t_{0}<t_{\fin}$, it can no longer be
the solution of the minimisation problem.

Taking into account \eqref{eq:tilde-k}, we have three different
regions, $A$, $B$ and $C$, for the optimal stiffness in the bounded
case $\kappa_{\bou}^{*}$:
\begin{equation}\label{eq:k*-bound}
 \kappa_{\bou}^{*}=
  \begin{cases}
    \begin{aligned}
    &  0, & & \text{if}
\quad \frac{\psi_{1}}
      {2\psi_{0}}y_{1} +1<0
      , & (A)
      \\ &
\frac{\psi_{1}} {2\psi_{0}\,y_{1}} +\frac{1}{y_{1}^{2}}, 
 & &\text{if}
\quad  0\leq\frac{\psi_{1}} {2\psi_{0}}y_{1}
+1\leq\frac{y_{1}^{2}}{\ym^{2}}
, &  (B) 
\\
& \kappa_{\max}, & &\text{if}
\quad \frac{\psi_{1}}{2\psi_{0}}y_{1}
+1>\frac{y_{1}^{2}}{\ym^{2}}
. &  (C)
    \end{aligned}
    \end{cases} 
\end{equation}
Along the same lines followed in the unbounded case, it is readily
shown that $y_{1,\bou}^{*}$ is linear in $t$ in region B with slope
$-\psi_{1}/(2\psi_{0})$, as predicted by \eqref{eq:canonical-unb}. We
denote this behaviour by $y_{1,\lin}(t)$. In appendix
\ref{app:bounded}, we show that if the system enters region A or
region C, it remains there. In other words, once the optimal solution
in region B ``touches'' the boundary at a certain time $t^{J}$,
i.e. $\kappa_{\bou}^{*}(t^{J})$ equals either $0$ or $\kappa_{\max}$,
it moves over the boundary from then on,
$\kappa_{\bou}^{*}(t)=\kappa_{\bou}^{*}(t^{J})$ for all $t>t^{J}$.  In
regions A and C, $y_{1}(t)$ is given by the solutions of
\eqref{eq:y-evol} corresponding to constant $\kappa=0$ and
$\kappa=\kappa_{\max}$, which we denote by $y_{1}(t)|_{\kappa=0}$ and
$y_{1}(t)|_{\kappa=\kappa_{\max}}$, respectively.

On physical grounds, we have three different cases depending on the values of the parameters:
namely $\{t_{\fin},\kappa_{\fin},\kappa_{\max}\}$.
\begin{enumerate}
\item\label{case1} The initial and final states cannot be linked in the given time
  $t_{\fin}$, which is too short. This is due to the impossibility of
  compressing (resp. decompressing) the system faster than with a STEP
  protocol with $\kappa(t)=\kappa_{\max}$ (resp. $\kappa(t)=0$).
\item\label{case2} The time interval $t_{\fin}$ is such that the
  connection is possible but not with the linear solution for the
  unbounded case $y_{1,\unb}^{*}(t)$, since the associated
  $\kappa^{*}_{\unb}(t) \notin [0,\kappa_{\max}]$ for a certain range
  of times inside $[0,t_{\fin}]$. In that case, we show below that the
  optimal protocol is built as a linear evolution of $y_{1}$ that
  matches continuously and smoothly (continuous first derivative) the
  solution of \eqref{eq:y-evol} with $\kappa=\kappa_{\max}$
  (resp. $\kappa=0$) in a compression (resp. decompression) process.
\item\label{case3} The given time $t_{\fin}$ is long enough to make
  the connection possible with the unbounded solution
  $y^{*}_{1,\unb}(t)$ because
  $\kappa_{\unb}^{*}(t)\in[0,\kappa_{\max}]$ for all times. In this
  case, the bounds do not affect the minimisation problem.
\end{enumerate}

\subsection{Decompression}\label{sec:decomp}

Let us look into the decompression case, in which $0\leq\kappa_{\fin}<1$. 
To begin with, we would like to discern when $\kappa_{\unb}^{*}(t)$
becomes negative. Looking at \eqref{eq:k-unb}, it is readily seen that
the first term on its rhs becomes smaller than the second one for
large enough $y_{1}$, and $y_{1,\unb}(t)$ increases linearly in
time. Therefore, the value of the final time $t_{\fin}$ below which
the unbounded solution ceases to be valid is determined by the
condition $\tilde{\kappa}_{\unb}(t_{\fin})=0$,
i.e. $y_{1,\fin}(y_{1,\fin}-1)=t_{\fin}$. Taking into account
\eqref{eq:bc-y}, this is equivalent to
$t_{\fin}=(1-\sqrt{\kappa_{\fin}})/\kappa_{\fin}$.

The condition $\kappa\geq 0$ implies that there are states that are
impossible to connect. The fastest decompression (shortest possible
$t_{\fin}$) corresponds to a STEP process, in which the stiffness is
instantaneously changed to $\kappa=0$ at $t=0^{+}$. In that case, we
have that $y_{1}(t)=\sqrt{1+2t}$ and thus
$y_{1,\fin}=\sqrt{1+2t_{\fin}}$. Recalling once more \eqref{eq:bc-y},
we conclude that the fastest decompression occurs for
$\kappa_{\fin}(1+2t_{\fin})=1$ or
$t_{\fin}=(1-\kappa_{\fin})/(2\kappa_{\fin})$.

Therefore, cases~\ref{case1}, \ref{case2} and \ref{case3} above
correspond here to:
\begin{enumerate}
\item Impossible to connect.
\begin{equation}\label{eq:tfmin-dec}
t_{\fin}< t_{d}^{\min}, \quad t_{d}^{\min}=\frac{1-\kappa_{\fin}}{2\kappa_{\fin}} 
.
\end{equation}
\item Matched solution, i.e. a first linear branch $y_{1,\lin}(t)$
  and a second branch moving over the line $\kappa=0$ of $S_{\bou}$,
  $y_{1}(t)|_{\kappa=0}$.
\begin{equation}\label{eq:tfu-dec}
 t_{d}^{\min}\leq t_{\fin}\leq t_{d}^{\unb}, \quad t_{d}^{\unb}=\frac{1-\sqrt{\kappa_{\fin}}}{\kappa_{\fin}}.
\end{equation}
\item Linear profile for the unbounded case  $y_{1,\unb}(t)$.
\begin{equation}
t_{\fin}> t_{d}^{\unb} 
.
\end{equation}
\end{enumerate}

In case \ref{case1}, there is no solution and we already know the
solution of case \ref{case3}. Then, we move on to solve case
\ref{case2}, for which the solution comprises two branches. First, a
branch corresponding to region B in \eqref{eq:k*-bound}, i.e. a linear
profile $y_{1,\lin}$ that verifies only the boundary condition at
$t=0$ and thus has one free parameter. This solution is valid in some
subinterval $[0,t_{d}^{J}]$, the free parameter can be considered to be
its constant slope $m_{d}=-\psi_{1}/(2\psi_{0})$, i.e.
\begin{equation}\label{eq:y1-lin-bou-dec}
  y_{1,\lin}(t)=1+m_{d} t, \quad t<t_{d}^{J}.
\end{equation}
Second, a branch corresponding to region A in \eqref{eq:k*-bound},
i.e. obtained by putting
$\kappa=0$ in \eqref{eq:y-evol}, $y_{1}(t)|_{\kappa=0}$. This branch
verifies the boundary condition at $t=t_{\fin}$ and is valid in the
complementary subinterval $[t_{d}^{J},t_{\fin}]$. Its specific form is
given by
\begin{equation}\label{eq:y1-kappa=0}
  y_{1}(t)|_{\kappa=0}=\sqrt{y_{1,\fin}^{2}-2(t_{\fin}-t)}, \quad t>t_{d}^{J}.
\end{equation}
Note that this branch does not contain any free parameter. The two
branches are matched at the joining time $t_{d}^{J}$ by imposing the
continuity of both $y_{1}(t)$ and $\dot{y}_{1}(t)$, i.e.
\begin{align}\label{eq:smoothness}
  y_{1,\lin}({t_{d}^{J}}^{-})=y_{1}|_{\kappa=0}({t_{d}^{J}}^{+}), & & \dot{y}_{1,\lin}({t_{d}^{J}}^{-})=\dot{y}_{1}|_{\kappa=0}({t_{d}^{J}}^{+}),
\end{align}
Note that this is consistent, any solution $y_{1}(t)$ of
\eqref{eq:y-evol} must be continuous for piecewise continuous
$\kappa(t)$. Moreover, since $\kappa(t)$ is continuous for the matched
solution at $t=t_{d}^{J}$,
$\kappa({t_{d}^{J}}^{-})=\kappa({t_{d}^{J}}^{+})=0$, $\dot{y}_{1}(t)$
must also be continuous there. We show in appendix \ref{app:bounded} that this
simplest approach is the correct one for our problem.

The continuity equations~\eqref{eq:smoothness} give rise to the conditions
\begin{subequations}\label{eq:cont-y-ydot}
\begin{align}
  y_{1,d}^{J}\equiv 1+m_{d}t_{d}^{J}&=\sqrt{y_{1,\fin}^{2}-2(t_{\fin}-t_{d}^{J})}, \label{eq:cont-y}\\
  m_{d}&=\frac{1}{\sqrt{y_{1,\fin}^{2}-2(t_{\fin}-t_{d}^{J})}}, \label{eq:cont-ydot}
\end{align}
\end{subequations}
which can be explicitly solved for $m_{d}$ and $t_{d}^{J}$, with the result
\begin{subequations}\label{eq:md-td}
\begin{align}
  t_{d}^{J}&=1+2t_{\fin}-y_{1,\fin}^{2}+\sqrt{1+2t_{\fin}-y_{1,\fin}^{2}}
  \\
  m_{d}&=\frac{1}
  {1+\sqrt{1+2t_{\fin}-y_{1,\fin}^{2}}}.
\end{align}
\end{subequations}
Note that the matching time $t_{d}^{J}$ is an increasing function of
$t_{\fin}$, vanishing in the limit as $t_{\fin}\to t_{d}^{\min}$ and
approaching $t_{\fin}$ in the limit as $t_{\fin}\to t_{d}^{\unb}$. In
fact, this solution only makes sense in case \ref{case2}: in case
\ref{case1}, the argument of the square root is negative whereas in
case \ref{case3} we have that $t_{d}^{J}>t_{\fin}$. Recall that
$y_{1,\fin}$ is given as a function of $\kappa_{\fin}$ by
\eqref{eq:bc-y}.

Then, the optimal protocol for the stiffness is
\begin{equation}\label{eq:kappa-d-sol}
  \kappa_{d}^{*}(t)=
    \begin{cases}
      \kappa_{\ini}=1, &  t=0, \\
       \displaystyle \frac{1}{[y_{1,\lin}(t)]^{2}}-\frac{m_{d}}{y_{1,\lin}(t)}, &
       0<t<t_{d}^{J}, \\
       0, & t_{d}^{J}\leq t<t_{\fin}, \\
      \kappa_{\fin}, &  t=t_{\fin}.
    \end{cases}
\end{equation}
The finite jumps of the stiffness at the initial and final times have
the same reason as in the unbounded case and thus we will not repeat
the discussion here. The initial jump in the stiffness decreases it to
a positive value, $\kappa_{d}^{*}(t=0^{+})=1-m_{d}$, and
$0<1-m_{d}<1$. In addition, note that $ \kappa_{d}^{*}(t)$ is continuous at
$t=t_{d}^{J}$, since the condition $m_{d}=1/y_{1,\lin}(t_{d}^{J})$
holds as a consequence of the continuity of the derivative of
$\dot{y}_{1}(t)$ at $t=t_{d}^{J}$, as expressed by
\eqref{eq:cont-ydot}. Consistently with our discussion below
\eqref{eq:md-td}, the expression in
\eqref{eq:kappa-d-sol} only makes sense for $t_{d}^{\min}\leq
t_{\fin}\leq t_{d}^{\unb}$.

\begin{figure}
\includegraphics[width=3.25in]{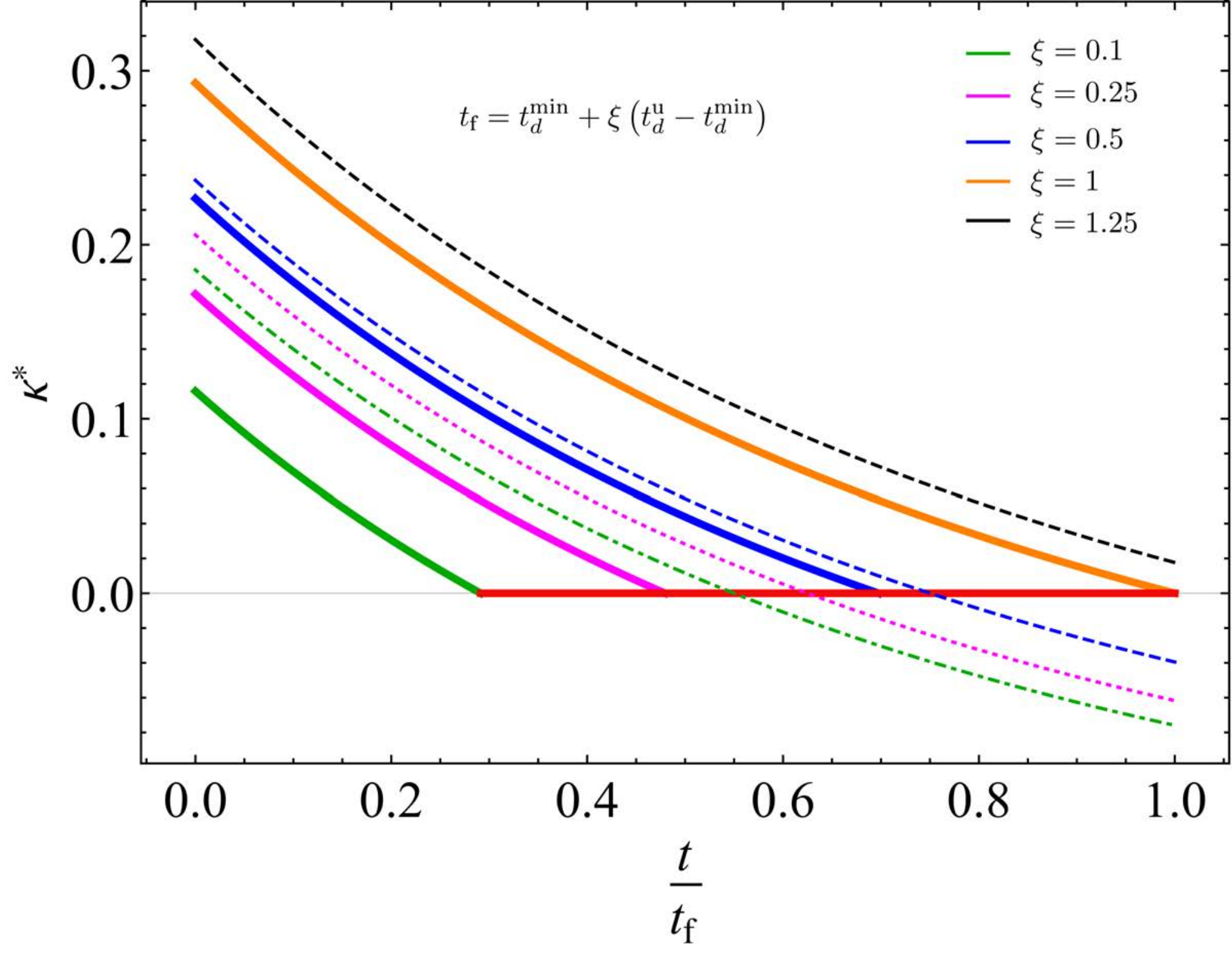}
\caption{\label{fig:sol-bounded-dec} Optimal protocols for the
  stiffness in the decompression case. We have chosen a decompression
  factor $\kappa_{\fin}=0.5$, for which the minimum time for
  connection is $t_{d}^{\min}=0.5$ and the time above which the
  unbounded solution works is $t_{d}^{\unb}=0.5858$. We compare the
  actual optimal protocol $\kappa_{d}^{*}(t)$ (thick solid) with the
  optimal protocol for unbounded stiffness $\kappa_{\unb}^{*}(t)$
  (dashed) for several values of the connection time
  $t_{\fin}=t_{d}^{\min}+\xi(t_{d}^{\unb}-t_{d}^{\min})$, where $\xi$
  from left to right is $0.1$ (green), $0.25$ (magenta), $0.5$ (blue),
  $1$ (orange) and $1.25$ (black). In order to show all the curves
  together, we plot them as a function of the scaled time
  $t/t_{\fin}$. The bounded solutions $\kappa_{d}^{*}(t)$ remain at the boundary $\kappa=0$ once they touch it at the corresponding
  matching time $t_{d}^{J}$.  For $\xi=1$ (orange curve), the solid and the dashed
  lines coincide, $t_{d}^{J}=t_{\fin}$. For the sake of clarity, the
  optimal protocols are shown for $t\in(0,t_{\fin})$; all of them have
  sudden jumps to $\kappa_{\ini}=1$ and $\kappa_{\fin}=0.5$ at $t=0$
  and $t=t_{\fin}$, respectively.}
\end{figure}

Optimal protocols for the decompression case are plotted in
Fig.~\ref{fig:sol-bounded-dec}. We have chosen $\kappa_{\fin}=0.5$ and
several values of the connection time $t_{\fin}$. The unbounded
solution $\kappa_{\unb}^{*}(t)$ (dashed lines) only works for the
longest time $t_{\fin}=t_{d}^{\unb}+0.25(t_{d}^{\unb}-t_{d}^{\min})$,
when it remains positive over the whole time interval. For the
remainder of shorter connecting times, $\kappa_{\unb}^{*}(t)$ becomes
negative as observed in the figure and the optimal protocol equals
$\kappa_{d}^{*}(t)$, as given by \eqref{eq:kappa-d-sol} (thick solid
lines). There is no solid line for the longest time, since
\eqref{eq:kappa-d-sol} is well-defined only for $t_{d}^{\min}\leq
t_{\fin}\leq t_{d}^{\unb}$.

\subsection{Compression}\label{sec:comp}

When the colloidal particle is compressed,
$\kappa_{\max}\geq \kappa_{\fin}>1$, the unbounded $\kappa_{u}^{*}(t)$ may
become greater than $\kappa_{\max}$. When this is the case, the
solution to the minimisation problem is built in a manner completely
analogous to the decompression case, but the second branch is obtained
by substituting $\kappa_{\max}$ into \eqref{eq:y-evol},
i.e. $y_{1}(t)|_{\kappa=\kappa_{\max}}$. Again, the two branches are
smoothly joined at a certain $t=t_{c}^{J}$, i.e. with $y_{1}$ and
$\dot{y}_{1}$ being continuous.

Since the scenario is analogous to that for decompression,  we do
not repeat the complete analysis here.
\begin{enumerate}
\item Impossible to connect:
\begin{equation}\label{eq:tfmin-comp}
t_{\fin}<t_{c}^{\min}, \quad t_{c}^{\min} = \frac{1}{2\kappa_{\max}} \ln \frac{\kappa_{\fin} \left(
    \kappa_{\max}-1 \right)}{\kappa_{\max}-\kappa_{\fin}},
\end{equation}
\item Matched solution, i.e. at first a linear branch $y_{1,\lin}(t)$
  and afterwards moving over the line $\kappa=\kappa_{\max}$ of $S_{\bou}$,
  $y_{1}(t)|_{\kappa=\kappa_{\max}}$,
\begin{equation}\label{eq:tfu-comp}
t_{c}^{\min}\leq t_{\fin}\leq t_{c}^{\unb}, \quad t_{c}^{\unb}=
\frac{ \sqrt{\kappa_{\fin}}-1}{\kappa_{\max}-\kappa_{\fin}},
\end{equation}
\item Linear profile for the unbounded case  $y_{1,\unb}(t)$.
\begin{equation}
t_{\fin} > t_{c}^{\unb} 
.
\end{equation}
\end{enumerate}

Again, we consider case \ref{case2}, for which the
solution comprises two branches. First, the linear branch $y_{1,\lin}$
valid in $[0,t_{c}^{J}]$, which
has the slope $m_{c}=-\psi_{1}/(2\psi_{0})$, 
\begin{equation}
  y_{1,\lin}(t)=1+m_{c} t, \quad t<t_{c}^{J}.
\end{equation}
Second, the branch obtained by substituting $\kappa=\kappa_{\max}$ in
\eqref{eq:y-evol}, $y_{1}(t)|_{\kappa=\kappa_{\max}}$, which verifies
the boundary condition at $t=t_{\fin}$ and is valid in
$[t_{c}^{J},t_{\fin}]$,
\begin{equation}
y_{1}(t)|_{\kappa=\kappa_{\max}}=\frac{\sqrt{1+(\kappa_{\max}y_{1,\fin}^{2}-1)e^{2\kappa_{\max}(t_{\fin}-t)}}}{\sqrt{\kappa_{\max}}}, \quad t>t_{c}^{J}.
\end{equation}
At the joining time $t_{c}^{J}$,
$y_{1}(t)$ and $\dot{y}_{1}(t)$ are continuous,
which yields
\begin{subequations}\label{eq:cont-y-ydot-comp}
\begin{align} y_{1,c}^{J}&\equiv 1+m_{c}t_{c}^{J}=\frac{\sqrt{1+(\kappa_{\max}y_{1,\fin}^{2}-1)e^{2\kappa_{\max}(t_{\fin}-t_{c}^{J})}}}{\sqrt{\kappa_{\max}}}, \label{eq:cont-y-comp}\\
  m_{c}&=- \frac{\sqrt{\kappa_{\max}}(\kappa_{\max}y_{1,\fin}^{2}-1)e^{2\kappa_{\max}(t_{\fin}-t_{c}^{J})}}{\sqrt{1+(\kappa_{\max}y_{1,\fin}^{2}-1)e^{2\kappa_{\max}(t_{\fin}-t_{c}^{J})}}}  . \label{eq:cont-ydot-comp}
\end{align}
\end{subequations}
At variance with the decompression case, this system cannot be
explicitly solved for $m_{c}$ and $t_{c}^{J}$ but we can obtain their
values for any set of the parameters
$\{t_{\fin},\kappa_{\fin},\kappa_{\max}\}$ numerically. Once more,
$y_{1,\fin}$ is given by \eqref{eq:bc-y} as a function of
$\kappa_{\fin}$. Note that $m_{c}<0$ because the standard deviation
decreases in time for compression.

Finally, we obtain the optimal protocol for the stiffness in the
compression process,
\begin{equation}\label{eq:kappa-c-sol}
  \kappa_{c}^{*}(t)=
    \begin{cases}
      \kappa_{\ini}=1, &  t=0, \\
       \displaystyle \frac{1}{[y_{1,\lin}(t)]^{2}}-\frac{m_{c}}{y_{1,\lin}(t)}, &
       0<t<t_{c}^{J}, \\
       \kappa_{\max}, & t_{c}^{J}\leq t<t_{\fin}, \\
      \kappa_{\fin}, &  t=t_{\fin}.
    \end{cases}
\end{equation}
Again, the initial jump in the stiffness goes in the ``right''
direction, it increases to $\kappa_{c}^{*}(t=0^{+})=1-m_{c}>1$ because $m_{c}$ as
given by \eqref{eq:cont-ydot-comp} is negative.

\begin{figure}
\includegraphics[width=3.25in]{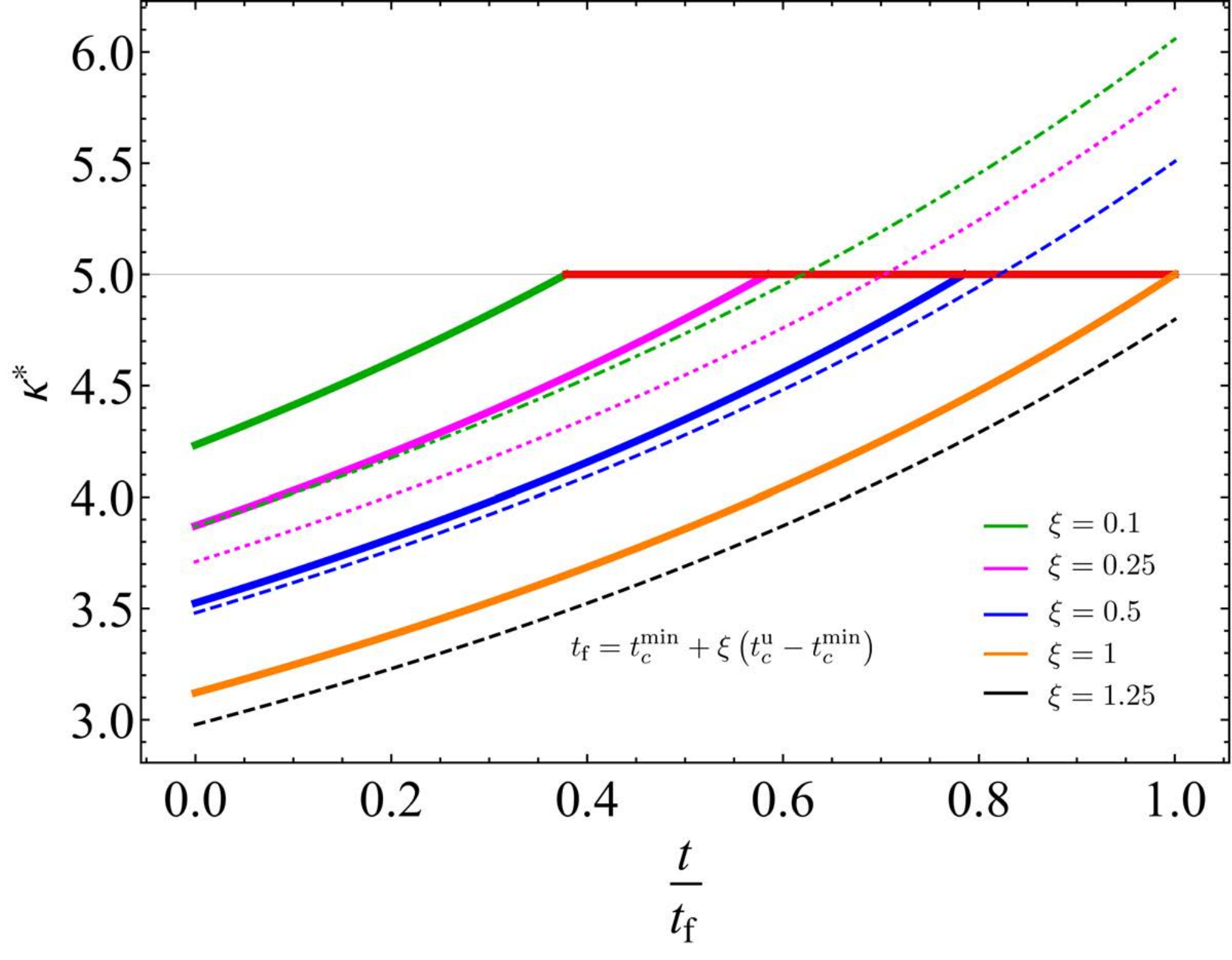}
\caption{\label{fig:sol-bounded-comp} Same as Fig. \ref{fig:sol-bounded-dec} for a compression. Here $\kappa_{\max}=5$
  (horizontal red thick line) and the compression factor is
  $\kappa_{\fin}=2$. With these parameters,
  $t_{c}^{\min}=0.09808$ and $t_{c}^{\unb}=0.1381$. We compare the
  actual optimal protocol $\kappa_{c}^{*}(t)$ (thick solid) with
  $\kappa_{\unb}^{*}(t)$ (dashed) for
  $t_{\fin}=t_{c}^{\min}+\xi(t_{c}^{\unb}-t_{c}^{\min})$, where $\xi$
  from left to right is $0.1$ (green), $0.25$ (magenta), $0.5$ (blue),
  $1$ (orange) and $1.25$ (black). As for decompression, the matched solutions $\kappa_{c}^{*}(t)$ remain at the
  boundary (here $\kappa=\kappa_{\max}$) for $t>t_{c}^{J}$, and the
  unbounded solution gives the correct optimal protocol only when
  $t_{\fin}>t_{c}^{\unb}$ because it remains smaller than
  $\kappa_{\max}$ for all times.}
\end{figure}

Figure~\ref{fig:sol-bounded-comp} is similar to
Fig.~\ref{fig:sol-bounded-dec} but for compression. We have
chosen the parameter values $\kappa_{\max}=5$ and
$\kappa_{\fin}=2$. The different curves correspond to different
connection times $t_{\fin}$. Similarly to the decompression case, the
optimal protocol $\kappa_{c}^{*}(t)\neq \kappa_{\unb}^{*}(t)$ except
for the longest time, since for the remainder of them
$\kappa_{\unb}^{*}(t)$ violates the inequality
$\kappa\leq \kappa_{\max}$. Also, the matching time $t_{c}^{J}$
increases with $t_{\fin}$, $t_{c}^{J}\to 0$ as
$t_{\fin}\to t_{c}^{\min}$ and $t_{c}^{J}\to t_{\fin}$ as
$t_{\fin}\to t_{c}^{\unb}$. Similarly to the decompression case,
\eqref{eq:cont-y-ydot-comp} and \eqref{eq:kappa-c-sol} only make sense
in case \ref{case2}.

\section{Phase diagram and average work}\label{sec:phase-diag}

\subsection{Inaccessible and accessible states}

Depending on the values of the parameters
$\{t_{\fin},\kappa_{\fin},\kappa_{\max}\}$, we have three different
``phases'' when the stiffness is bounded, which correspond to each of
the cases enumerated in the previous section. For each value of the
maximum stiffness $\kappa_{\max}$, there are target points
$(\kappa_{\fin},t_{\fin})$ that
\begin{enumerate}
\item are inaccessible; there is no
control $\kappa(t)$ capable of linking the initial and final states,
\item
can be reached by means of a matched solution; the optimal
control moves partially over the boundary of the rectangle
$S_{\bou}\equiv [0,t_{\fin}]\times[0,\kappa_{\max}]$, and
\item
can be
reached with the optimal control for the unbounded case;
the standard deviation has the simple linear form $y_{1,\unb}(t)$
in \eqref{eq:y1-linear}.
\end{enumerate}

In order to look into the different phases, it is worth going to the
natural time scale for relaxation at the final stiffness
$\kappa_{\fin}$, i.e we define
\begin{equation}\label{eq:tau-def}
\tau=\kappa_{\fin}t.
\end{equation}
In this time scale, the equilibration time is the same for all
$\kappa_{\fin}$, $\tau_{\eq}\simeq 3$, see \eqref{eq:teq}. Therefore,
the value of the connection time in the $\tau$ scale,
$\tau_{\fin}=\kappa_{\fin}t_{\fin}$, directly gives the acceleration
of the ESE process with respect to the STEP one. The times separating
the different regions (inaccessible, matched, unbounded) are readily
obtained from \eqref{eq:tfmin-dec} and \eqref{eq:tfu-dec} in
decompression
\begin{equation}\label{eq:tau-crit-dec}
\tau_{d}^{\min}=\frac{1-\kappa_{\fin}}{2}, \quad
\tau_{d}^{\unb}=1-\sqrt{\kappa_{\fin}},
\end{equation}
and
\eqref{eq:tfmin-comp} and \eqref{eq:tfu-comp} in compression,
\begin{equation}\label{eq:tau-crit-comp}
\tau_{c}^{\min}=\frac{\kappa_{\fin}}{2\kappa_{\max}} \ln \frac{\kappa_{\fin} \left(
    \kappa_{\max}-1 \right)}{\kappa_{\max}-\kappa_{\fin}}, \quad 
\tau_{c}^{\unb}=\frac{\kappa_{\fin}\, (\sqrt{\kappa_{\fin}}-1)}{\kappa_{\max}-\kappa_{\fin}}
\end{equation}

\begin{figure}
\includegraphics[width=3.25in]{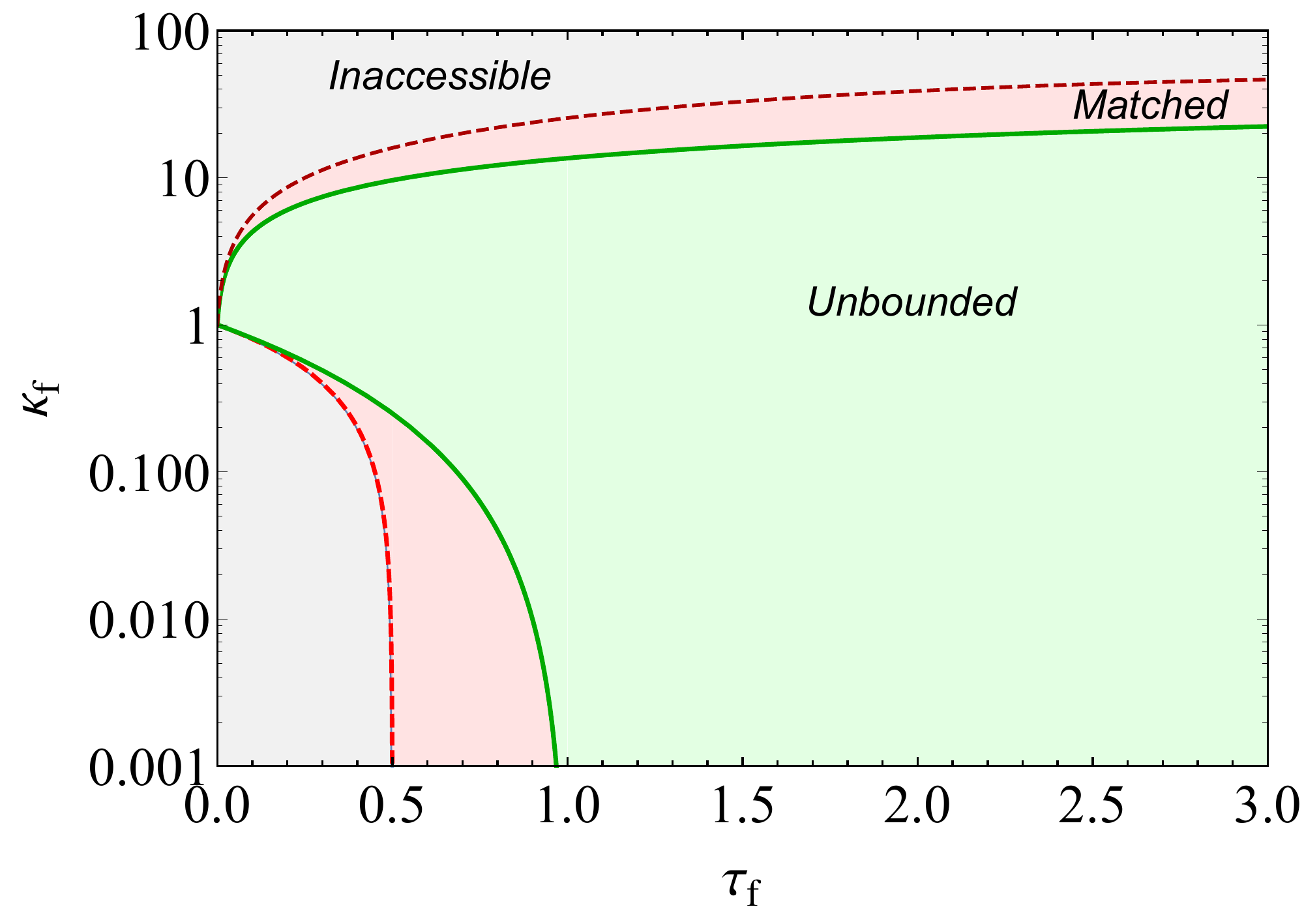}
  \caption{\label{fig:phase-diag} Phase diagram of the system in the
    $(\tau_{\fin},\kappa_{\fin})$ plane for the control problem with
    bounded $\kappa$, $0<\kappa<\kappa_{\max}$. Specifically, we are
    showing the case $\kappa_{\max}=50$. Note the logarithmic scale on
    the vertical axis. Target points
    $(\tau_{\fin},\kappa_{\fin})$ inside the grey regions cannot be
    reached. As compared to the unbounded case, on the one hand, the
    solution of the optimal control problem remains unchanged for
    target points inside the green region, since they can be reached by
    the optimal control for the unbounded case
    $\kappa_{\unb}^{*}(\tau)$. On the other hand, target points inside
    the red regions cannot be reached with the unbounded
    solution. Thus, there appears a new optimal solution
    $\kappa_{\bou}^{*}(\tau)$, which comprises two branches that are
    smoothly matched, as described in Sec.~\ref{sec:bounded}.
}
\end{figure}

In Fig.~\ref{fig:phase-diag}, we plot the different regions in the
plane $(\tau_{\fin},\kappa_{\fin})$ for the specific case
$\kappa_{\max}=50$. We have shaded regions in (i) grey, (ii) red and
(iii) green. The dashed red lines separate regions (i) and (ii),
i.e. they are given by $\tau_{\fin}=\tau_{p}^{\min}$, where $p=c$ or
$d$ depending on the type of process, compression or
decompression. The optimal protocol over these lines is an initial
abrupt change from $\kappa(0)=\kappa_{\ini}=1$ to $\kappa=0$ in the
decompression process (to $\kappa=\kappa_{\max}$ for compression) and
another sudden jump from this value to the target stiffness
$\kappa_{\fin}$ at the final time. The solid green lines separate
regions (ii) and (iii), i.e. they are given by
$\tau_{\fin}=\tau_{p}^{\unb}$, again $p=c$ or $d$ depending on the
type of process. Over these lines, the unbounded solution becomes
valid throughout the whole time interval, reaching the border
($\kappa=0$ for decompression, $\kappa=\kappa_{\max}$ for compression) at $t=t_{\text{\fin}}^{-}$.

The decompression case deserves further commenting. Both the minimum
connection time $\tau_{d}^{\min}$ and the time above which the
unbounded solution is valid $\tau_{d}^{\unb}$ are bounded, specifically
\begin{equation}
\tau_{d}^{\min}\leq \tau_{d}^{(1)}=1/2, \quad \tau_{d}^{\unb}\leq\tau_{d}^{(2)}=1.
\end{equation}
As it is clearly seen in Fig.~\ref{fig:phase-diag}, this means that
$\tau_{d}^{(2)}=1$ is a critical time in decompression: above it, the
initial equilibrium state can always be connected to another
equilibrium state corresponding to an arbitrary value of the stiffness
with the protocol valid for the unbounded case. Moreover,
$\tau_{d}^{(1)}=1/2$ is a second critical time in decompression: for
$\tau_{d}^{(1)}<\tau_{\fin}<\tau_{d}^{(2)}$, all the final equilibrium
states are accessible but the unbounded solution is only valid for
weak enough decompression---meaning $\kappa_\fin$ smaller but not too
far from unity, whereas for $\tau_{\fin}<\tau_{d}^{(1)}$ there appear
inaccessible states.  This is to be contrasted with the compression
case. In this latter case, the three possible phases, inaccessible,
matched and unbounded, are possible for all the connecting times
$\tau_{\fin}$.

Note that the existence of an upper (resp.~lower) bound on
$\kappa$ does not affect the decompression (resp.~compression)
case. This stems from the monotonicity of the optimal protocols for
the stiffness, as explicitly proven in Appendix \ref{app:bounded}.

\subsection{Properties of the mean work}

At this point, it is worth looking into the optimal average work and
elucidate how the problem changes upon constraining the stiffness $\kappa$. The optimal value
for the irreversible work $\calW^{*}$ can be computed in regions (ii)
and (iii), when the connection between the initial and final states is
possible. In region (iii), the bound on the stiffness plays no role
for calculating $\calW_{\irr}^{*}$,
$\calW_{\irr}^{*}=\calW_{\irr,\unb}^{*}$ as given by
\eqref{eq:opt-work-unb}. In region (ii), we have to use the matched
solutions in Secs.~\ref{sec:decomp} and \ref{sec:comp} to derive the
minimum work. We employ again $p=c$ or $d$ to label the kind of
process. The integral in \eqref{eq:functional-def} is split into two
parts: the first one from $0$ to $t_{p}^{J}$, where $y_{1}(t)$ is
linear in time with slope $m_{p}$, and the second one from $t_{p}^{J}$
to $t_{\fin}$, where $y_{1}(t)$ is given by the boundary solution
$y_{1}(t)|_{\kappa=\kappa_{p}}$; $\kappa_{p}$ stands for the relevant
boundary value of $\kappa$, $\kappa_{d}=0$ and
$\kappa_{c}=\kappa_{\max}$. Then,
\begin{equation}
\calW_{\irr,p}^{*}=m_{p}^{2}t_{p}^{J}+\int_{t_{p}^{J}}^{t_{\fin}}dt\, \left[\dot{y}_{1}(t)|_{\kappa=\kappa_{p}}\right]^{2},
\end{equation}
Integrating over $y_{1}$ instead of $t$ in the second term of the rhs
and making use of \eqref{eq:f1-def} and the continuity of $y_{1}$ at
the matching time,
$y_{1,p}(t_{p}^{J-})=y_{1,p}(t_{p}^{J+})=y_{1,p}^{J}$, one gets
\begin{equation}\label{eq:Wirr-p-*}
  \calW_{\irr,p}^{*}=m_{p}^{2}t_{p}^{J}+\ln
  \frac{y_{1,\fin}}{y_{1,p}^{J}}-\kappa_{p}\frac{y_{1,\fin}^{2}-(y_{1,p}^{J})^{2}}{2}.
\end{equation}

Let us particularise \eqref{eq:Wirr-p-*} for decompression and
compression. First, in the decompression case we have that
\begin{equation}\label{eq:irr-work-opt-bou-decomp}
  \calW_{\irr,d}^{*}=m_{d}^{2}t_{d}^{J}+\ln\frac{y_{1,\fin}}{y_{1,d}^{J}},
\end{equation}
in which $m_{d}$ and $t_{d}^{J}$ are given by \eqref{eq:md-td} in terms of
$(t_{\fin},\kappa_{\fin})$, and $y_{1,d}^{J}$ is the value of $y_{1}$
at the joining time $t_{d}^{J}$ as defined in \eqref{eq:cont-y}.
Second, for compression we obtain
\begin{equation}\label{eq:irr-work-opt-bou-comp}
  \calW_{\irr,c}^{*}=m_{c}^{2}t_{c}^{J}+\kappa_{\max}\frac{
    \left(y_{1,c}^{J}\right)^{2}-y_{1,\fin}^{2}}{2}-
\ln\frac{y_{1,c}^{J}}{y_{1,\fin}},
\end{equation}
where $m_{c}$ and $t_{c}^{J}$ are the solutions of the system of
equations \eqref{eq:cont-ydot-comp}, and $y_{1,c}^{J}$ is given by
\eqref{eq:cont-y-comp}.

\begin{figure}
  \includegraphics[width=3.25in]{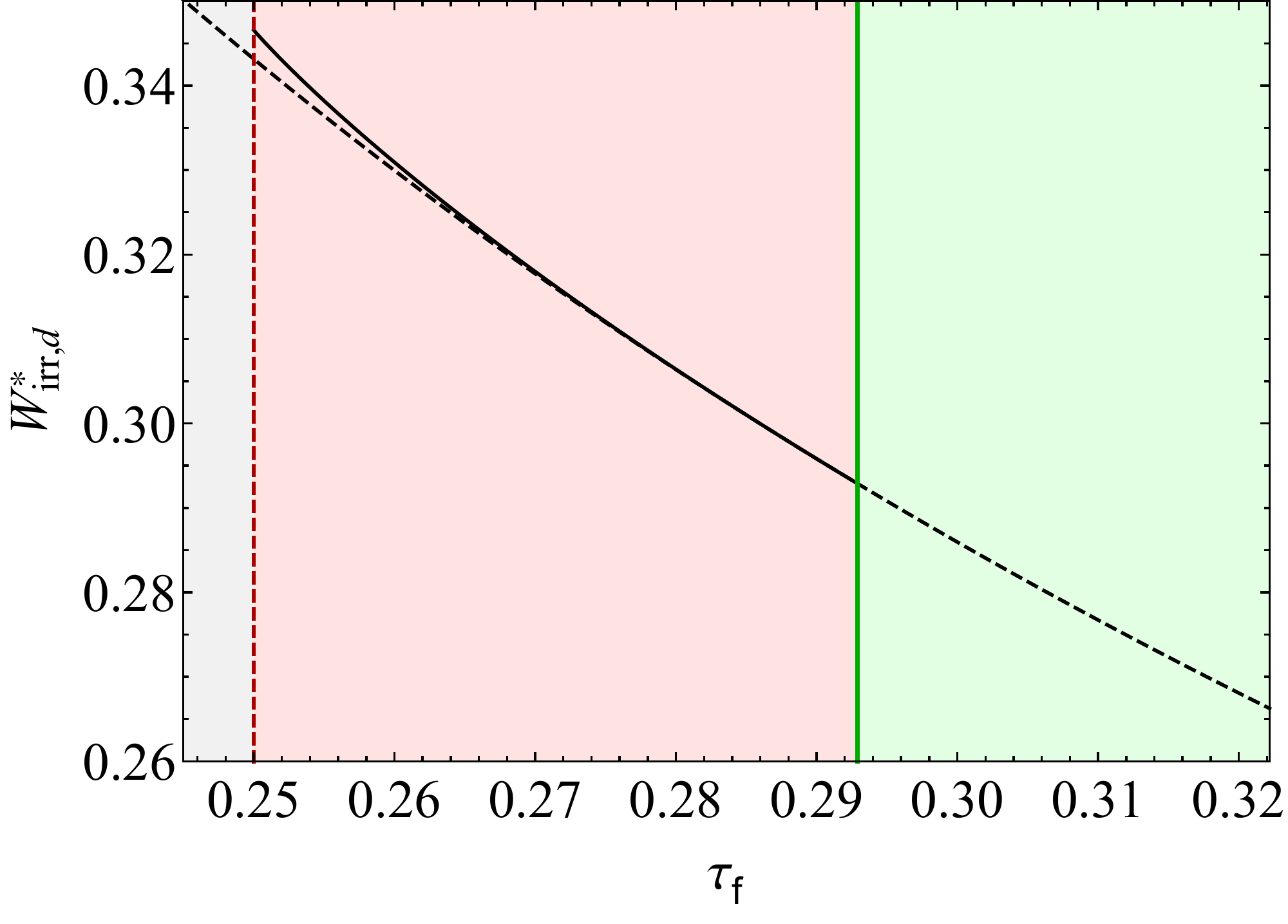}
  \includegraphics[width=3.25in]{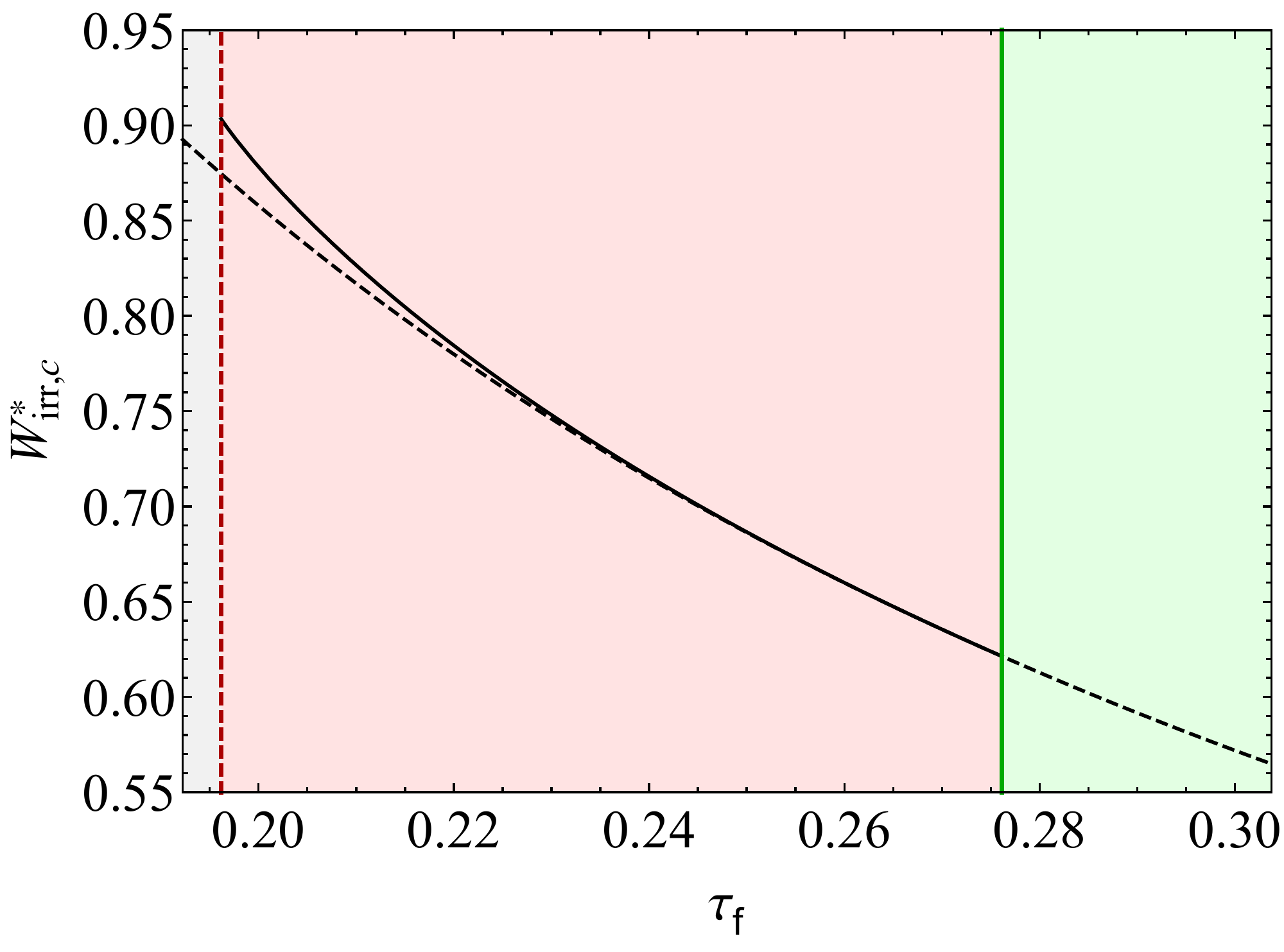}
  \caption{\label{fig:work-func-time} Optimal irreversible work of the
    system as a function of the final time
    $\tau_{\fin}=\kappa_{\fin} t_{\fin}$. Top: decompression
    ($\kappa_{\fin}=0.5$). Bottom: compression ($\kappa_{\fin}=2$ with
    $\kappa_{\max}=5$). Dashed lines correspond to the unconstrained
    result \eqref{eq:opt-work-unb}, whereas the solid lines stand for
    the solutions in the constrained case. Note that the latter only
    exist within the red region---colour code for the regions is the
    same as in Fig.~\ref{fig:phase-diag}, and are given by
    \eqref{eq:irr-work-opt-bou-decomp} (decompression) and
    \eqref{eq:irr-work-opt-bou-comp} (compression). The minimum
    irreversible work for the bounded case is, logically, always above
    that for the unbounded situation.  }
\end{figure}

In what follows, we plot with dashed lines the optimal work coming from
the unbounded expression, as given by \eqref{eq:opt-work-unb}.  Solid
lines are used for the optimal work when the bound
$0\leq \kappa\leq\kappa_{\max}$ is relevant,
\eqref{eq:irr-work-opt-bou-decomp} for decompression and
\eqref{eq:irr-work-opt-bou-comp} for compression. In addition, we
have shaded the different regions with the same colour code employed in
the phase diagram. The solid lines are always above the dashed ones,
because the minimum with no constraints is logically lower than the
constrained one.

First, we investigate the optimal work as a function of the final time
$\tau_{\fin}$, for different values of $\kappa_{\fin}$. Specifically,
we consider a compression protocol with $\kappa_{\fin}=2$ and a
decompression protocol with $\kappa_{\fin}=0.5$ in
Fig.~\ref{fig:work-func-time}. The stiffness is bounded in the
interval $0\leq \kappa \leq \kappa_{\max}=5$. The difference between
the constrained and unconstrained optimal values of the work becomes
more important as the connection time $\tau_{\fin}$ decreases, as
discussed below.

\begin{figure}
  \includegraphics[width=3.25in]{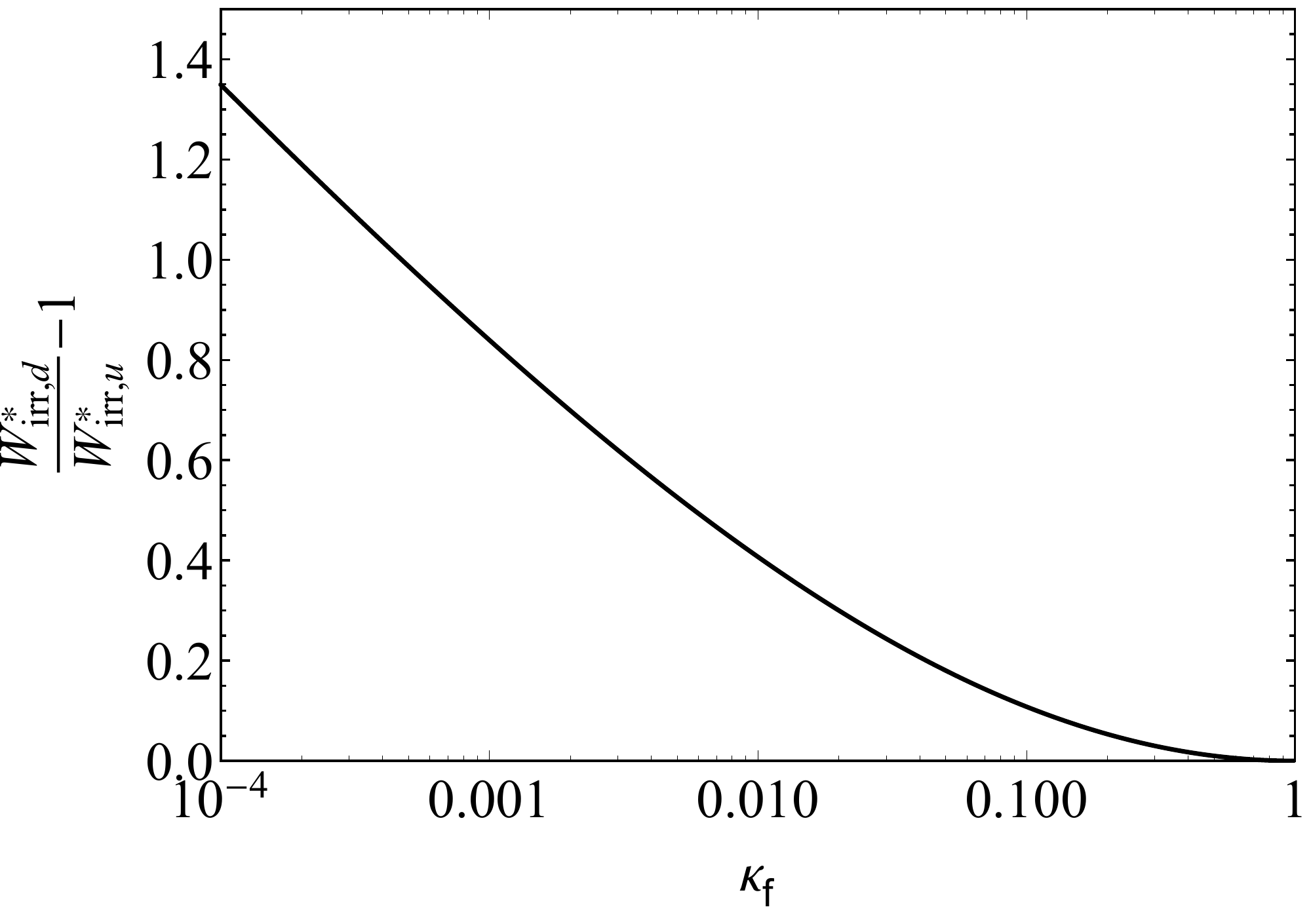} 
  \caption{\label{fig:work-irr-comparison} Relative difference between
    the bounded and unbounded optimal values of the irreversible work
    as a function of $\kappa_{\fin}$. The plot corresponds to the
    decompression region $\kappa_{\fin}<1$. Specifically, we use the
    work values at the minimum connection time $\tau_{d}^{\min}$, for
    which the relative difference attains its largest value. Note the
    divergence that appears in the limit as $\kappa_{\fin}\to 0$.  }
\end{figure}

Let us investigate the decompression case in more detail.  We
focus on the difference between the actual optimal work
$\calW_{\irr,d}^{*}$ and its value for the unconstrained case
$\calW_{\irr,\unb}^{*}$ for the minimum connection time
$\tau_{\fin}\to \tau_{d}^{\min}$ (or $t_{\fin}\to t_{d}^{\min}$),
which is given by \eqref{eq:tau-crit-dec}. At this point, this
difference reaches its maximum value. Therefore, the first term on the
rhs of \eqref{eq:irr-work-opt-bou-decomp} for the optimal work
$\calW_{\irr,d}^{*}$ does not contribute thereto, because
$t_{d}^{J}\to 0$, and we have
\begin{equation}\label{eq:Wirr-limits}
  \lim_{t_{\fin}\to
    t_{d}^{\min}}\calW_{\irr,d}^{*}=-\frac{1}{2}\ln\kappa_{\fin},
  \quad \lim_{t_{\fin}\to
    t_{d}^{\min}}\calW_{\irr,\unb}^{*}=2\frac{1-\sqrt{\kappa_{\fin}}}
  {1+\sqrt{\kappa_{\fin}}}.
\end{equation}
In Fig.~\ref{fig:work-irr-comparison} we plot the relative difference
$(\calW_{\irr,d}^{*}-\calW_{\irr,\unb}^{*})/\calW_{\irr,\unb}^{*}$ as
a function of $\kappa_{\fin}$. It remains small for
$\kappa_{\fin}\gtrsim 0.3$, for instance for $\kappa_{\fin}=0.5$ it is
below $1\%$. As $\kappa_{\fin}$ decreases it starts to grow; in fact,
$\calW_{\irr,d}^{*}$ diverges as $\kappa_{\fin}\to 0$. For example, for
$\kappa_{\fin}=0.1$ the relative difference is around $10\%$, for
$\kappa_{\fin}=0.01$ it has increased to $40\%$ and for
$\kappa_{\fin}=10^{-3}$ it exceeds $80\%$. \\

\begin{figure}
  \includegraphics[width=3.25in]{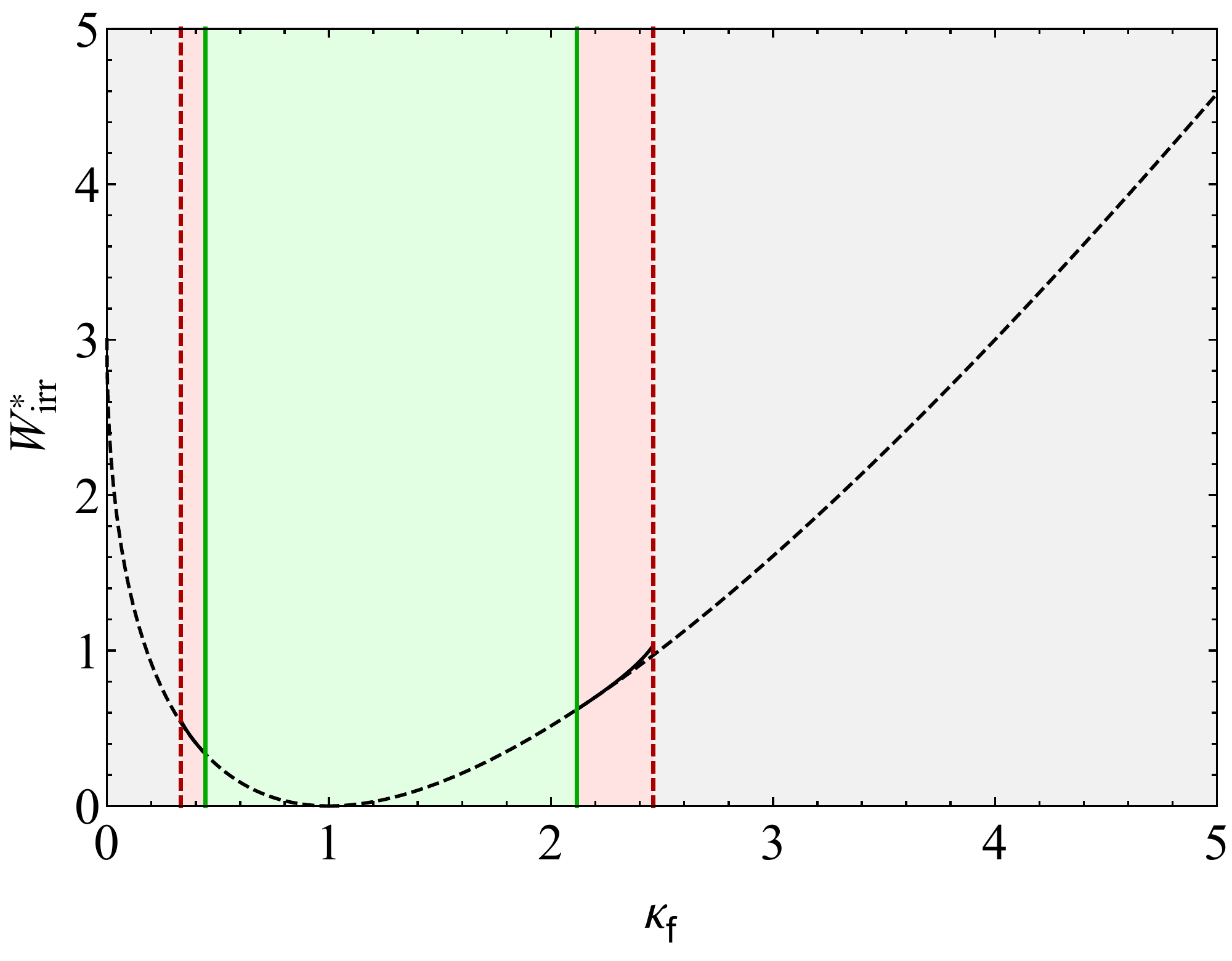} \includegraphics[width=3.25in]{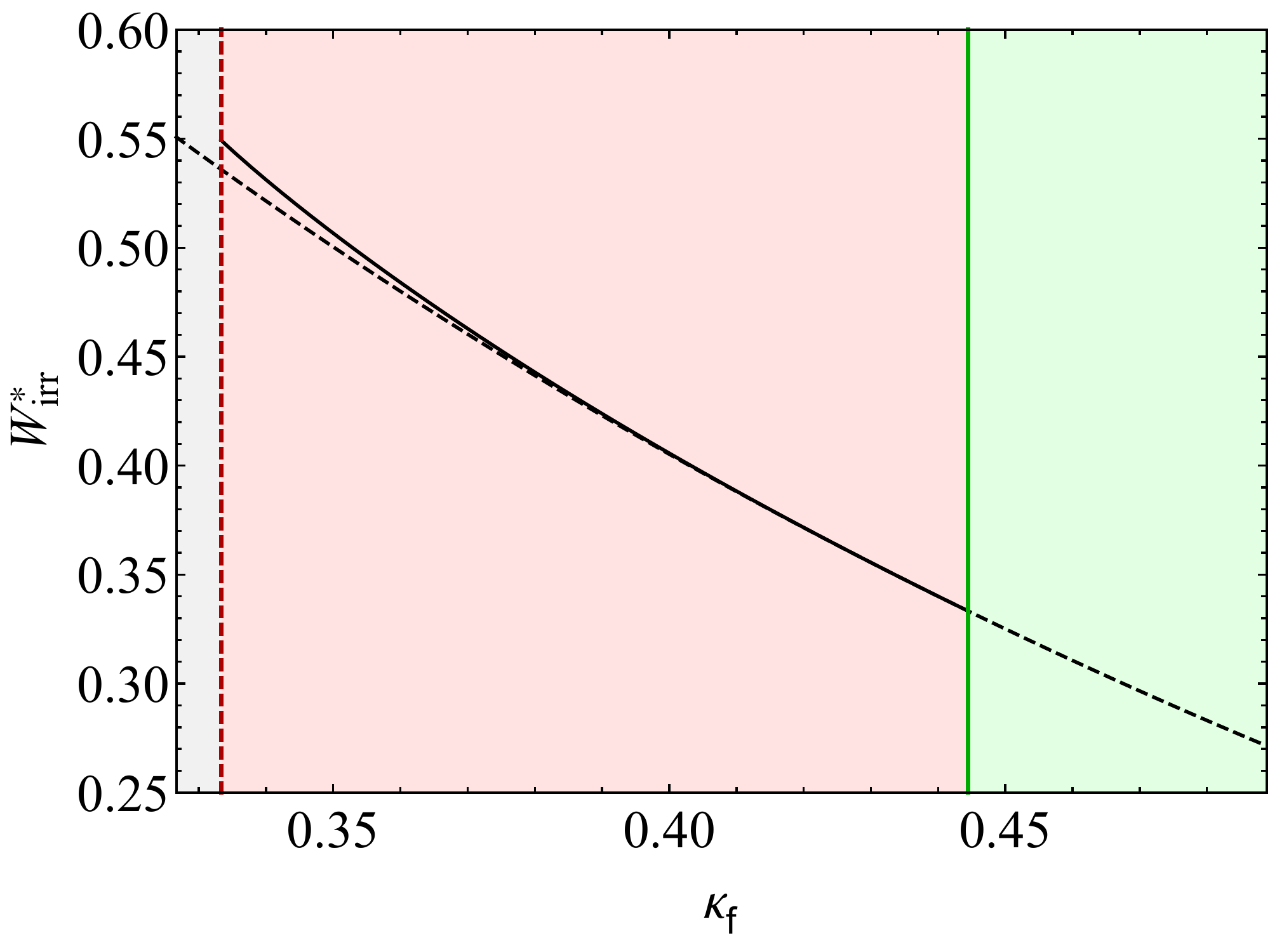}
  \caption{\label{fig:work-func-kappa} (Top) Optimal irreversible work
    of the system as a function of the compression ratio
    $\kappa_{\fin}$. The graph correspond to the parameter values
    $\kappa_{\max}=5$ and $\tau_{\fin}=1/3$, so that the
    connection time is roughly one tenth of the equilibration time
    $\tau_{\eq}\simeq 3$.  Colour code is the same as in
 Fig.~\ref{fig:work-func-time}.
    (Bottom) Zoom into the red region (matched solution) for
    decompression. }
\end{figure}

Second, we study the optimal work as a function of the
\textit{compression ratio} $\kappa_{\fin}$ for a fixed value of the
connection time $\tau_{\fin}$.
Similarly to the situation found when $\tau_{\fin}$ was varied for fixed $\kappa_{\fin}$, we have again inaccessible, matched
and unbounded regions. Specially interesting
is the decompression case, in principle the minimum value of the
stiffness $\kappa_{d}^{\min}$ for having connected states and the
value $\kappa_{d}^{\unb}$ above which the unbounded solution works
should be obtained by using
\eqref{eq:tau-crit-dec}. Notwithstanding, the situation is a little
more complex. Specifically we have that
\begin{equation}
\kappa_{d}^{\min}=\begin{cases}
1-2\tau_{\fin}, & \tau_{\fin}\leq \tau_{d}^{(1)}, \\
0, & \tau_{\fin}>\tau_{d}^{(1)}.
\end{cases} \; 
\sqrt{\kappa_{d}^{\unb}}=
\begin{cases} 1-\tau_{\fin}, & \tau_{\fin}\leq \tau_{d}^{(2)}, \\
0, & \tau_{\fin}>\tau_{d}^{(2)}.
\end{cases}
\end{equation}
The piecewise definitions of $\kappa_{d}^{\min}$ and
$\kappa_{d}^{\unb}$ are readily rationalised by looking at
Fig.~\ref{fig:phase-diag}: obtaining $\kappa_{d}^{\min}$
\eqref{eq:tau-crit-dec} only makes sense as long as
$\tau_{\fin}\leq\tau_{d}^{(1)}=1/2$, above it $\kappa_{d}^{\min}=0$
because all the states with $\tau_{\fin}>\tau_{d}^{(1)}$ are
accessible. A similar reasoning applies to $\kappa_{d}^{\unb}$: for
$\tau_{\fin}>1$, all the states can be connected with the unbounded
solution.  The most interesting region in the ESE context is
$\tau_{\fin}<\tau_{d}^{(1)}$, which corresponds to the higher
acceleration of the equilibration process,
$\tau_{\fin}/\tau_{\eq}<1/6$.

Figure~\ref{fig:work-func-kappa} corresponds to the specific case
$\kappa_{\max}=5$ and $\tau_{\fin}=1/3$. Therefore, the connection
time is roughly one tenth of the equilibration time,
$\tau_{\fin}/\tau_{\eq}=1/9$, and we plot compression
($\kappa_{\fin}>1$) and decompression ($\kappa_{\fin}<1$) processes in
the same graph.  For these values of the parameters, the main effect
of the bounds is the reduction of the effectively accessible region
for $\kappa$, which is much smaller than the whole interval
$[0,\kappa_{\max}]$.  The matched solutions are needed in two layers
close to the borders of the accessible region, but the differences
between the bounded optimal work and the unbounded value are quite
moderate. 

\begin{figure}
  \includegraphics[width=3.25in]{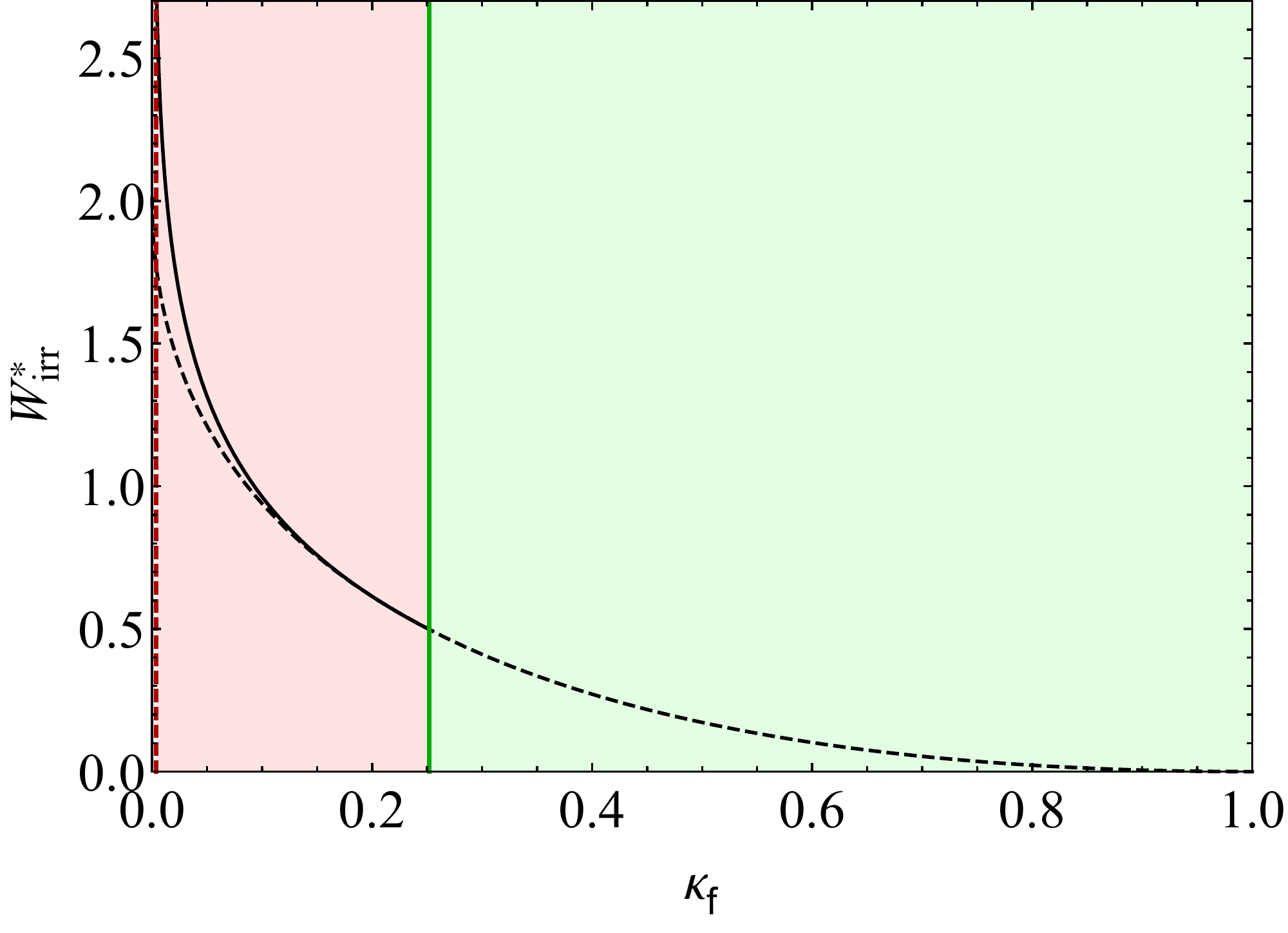} \includegraphics[width=3.25in]{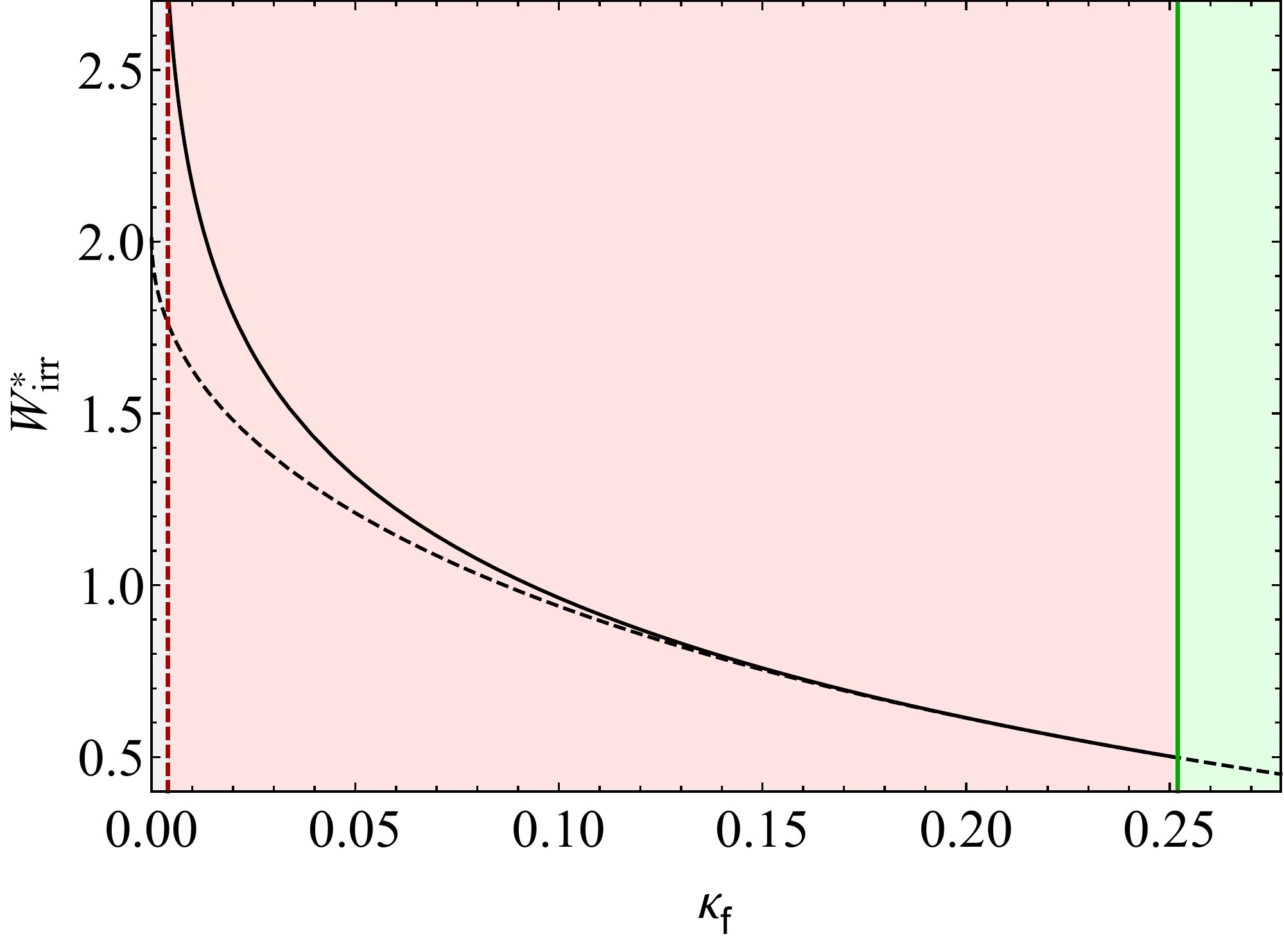}
  \caption{\label{fig:work-f-kappa-critical} (Top) Optimal irreversible work
    as a function of $\kappa_{\fin}$ for a connecting time close to the critical
    value $\tau_{d}^{(1)}=1/2$. We show only the decompression region
    $\kappa_{\fin}<1$ for the specific case
    $\tau_{\fin}=0.498$.
    (Bottom) Zoom into the red region, inside which the matched solution
    is needed. It is observed that the differences between the
    bounded and unbounded optimal values can become quite large, up to
    the order of $60\%$ in this particular case.
  }
\end{figure}

\begin{figure}
\includegraphics[width=3.25in]{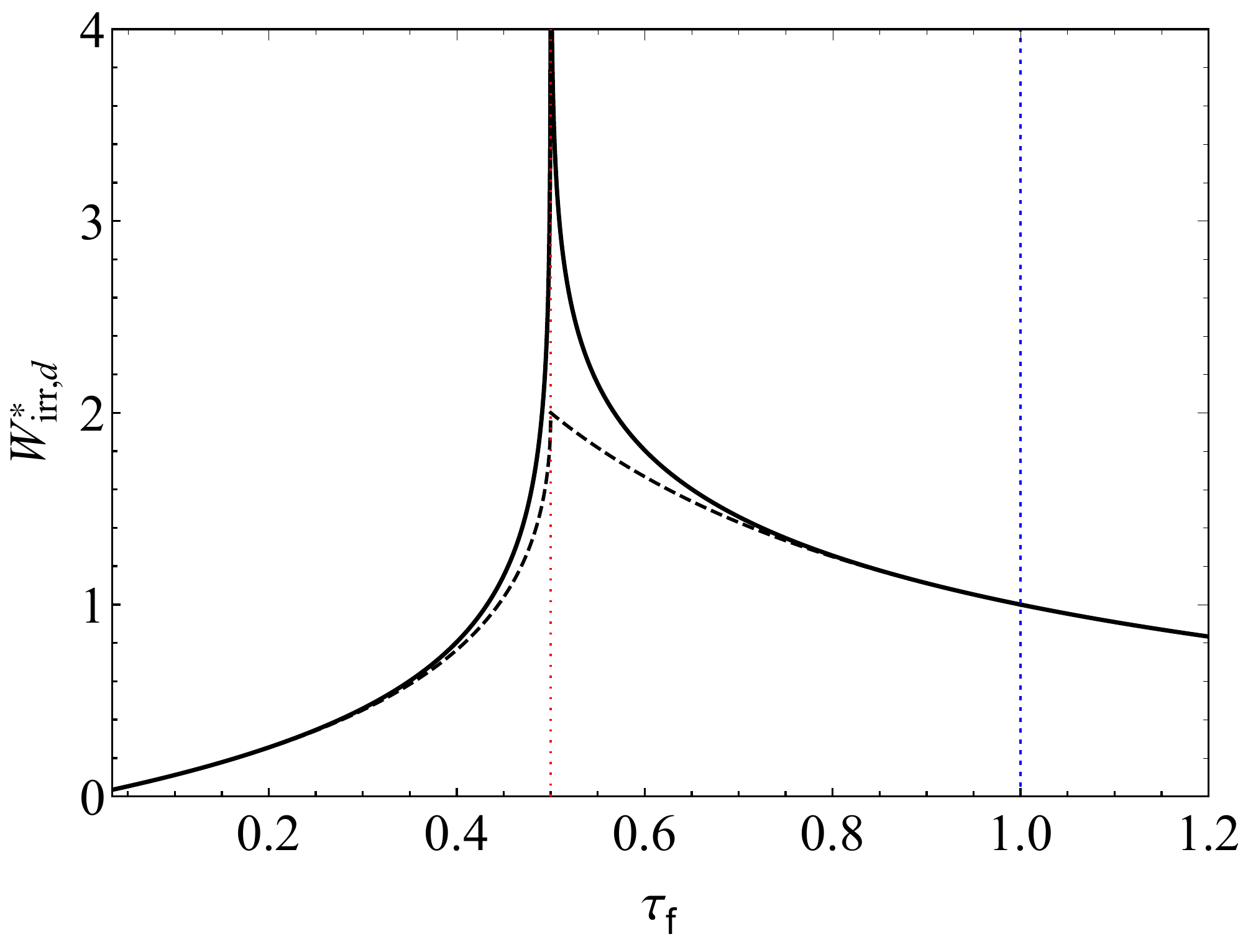}
\caption{\label{fig:work-irr-comparison-tauf} Comparison between the
  bounded and unbounded optimal values of the irreversible work as a
  function of $\tau_{\fin}$. Specifically, the irreversible work is
  evaluated at the minimum value of the stiffness allowing for
  connection $\kappa_{d}^{\min}$, always in the decompression
  region. The divergence of the bounded optimal work
  $\calW_{\irr,d}^{*}$ (solid line) at the critical time
  $\tau_{d}^{(1)}=1/2$ (red dotted line) is clearly observed,
  whereas $\calW_{\irr,u}^{*}$ (dashed line) remains finite
  throughout. For $\tau_{\fin}$ longer than $\tau_{d}^{(2)}=1$ (blue
  dotted line), $\calW_{\irr,d}^{*}$ and $\calW_{\irr,\unb}^{*}$ are
  identical, as discussed in the text.  }
\end{figure}

We consider a value of the connecting time close to the critical value
$\tau_{d}^{(1)}$ in Fig.~\ref{fig:work-f-kappa-critical},
specifically $\tau_{\fin}=0.498$. The inaccessible region becomes very
small, since $\kappa_{d}^{\min}=0.004$ but the bounded irreversible
work $\calW_{\irr,d}^{*}$ is about $60\%$ higher than the unbounded
irreversible value $\calW^{*}_{\irr,\unb}$ at
$\kappa_{\fin}=\kappa_{d}^{\min}$. In fact, as
$\tau_{\fin}\to\tau_{d}^{(1)}$ we have that
$\kappa_{d}^{\min}\to 0$ and the corresponding $\calW_{\irr,d}^{*}$
diverges logarithmically whereas the bounded value remains finite,
$\calW_{\irr,\unb}^{*}\to 2$, as expressed by \eqref{eq:Wirr-limits}.
We further illustrate this fact by plotting both $\calW_{\irr,d}^{*}$
and $\calW_{\irr,\unb}^{*}$ at $\kappa_{\fin}=\kappa_{d}^{\min}$ as a
function of $\tau_{\fin}$ in
Fig.~\ref{fig:work-irr-comparison-tauf}. It is observed that,
consistently with the discussion above and the picture shown in
Fig.~\ref{fig:work-irr-comparison}, the difference between the two are
largest for $\tau_{\fin}\to\tau_{d}^{(1)}$. In principle, it may
seem strange that $\calW_{\irr,d}^{*}$ and $\calW^{*}_{\irr,\unb}$ tend to
coincide in the limit as $\tau_{\fin}\to 0$. Looking once more at
Fig.~\ref{fig:phase-diag}, it is seen that the inaccessible region
grows as $\tau_{\fin}$ is decreased and fills the whole decompression
region as $\tau_{\fin}\to 0$, i.e.  $\kappa_{d}^{\min}\to 1$ in this
limit. Therefore, a very large acceleration of the process is only
possible in the \textit{linear response regime}
$1-\kappa_{\fin}\ll 1$, for which both works are infinitesimally
small. In fact, their relative difference can also be shown to be very
small.

\section{Conclusions}\label{sec:conclusions}

In experiments with confined colloids, a natural constraint on the
trap stiffness is that expressed by \eqref{eq:non-holonomic}. This
non-holonomic constraint makes it impossible to solve the minimisation
problem of the work by employing the usual approach involving the
Euler-Lagrange equations. Instead, it is necessary to address the
problem by employing the tools of control theory, specifically
Pontryagin's maximum principle. Interestingly, a similar approach
based on control theory has been recently applied to address the
minimisation of entropy production in the trapped colloidal particle
problem~\cite{muratore-ginanneschi_application_2017}, but with
``bounded accelerations''. The relevance of these bounds, which were
originally introduced to regularise the jumps of the stiffness at the
initial and final times \cite{aurell_boundary_2012}, for experiments
is not obvious.

The bounds on the stiffness strongly modify the problem of minimising
the work performed on the colloidal particle. The solution for
unbounded stiffness, in which the standard deviation $y_{1}$ connects
linearly the initial and final states, is no longer valid in general:
the associated optimal stiffness $\kappa_{\unb}^{*}(t)$ violates the
inequality \eqref{eq:non-holonomic} for short enough connecting times
$t_{\fin}$. First, there appear minimum times for connecting the
initial and final states, since it is impossible to compress
(resp. decompress) the system with any control $\kappa(t)$ faster than with
the one corresponding to $\kappa(t)=\kappa_{\max}$
(resp. $\kappa(t)=0$) for all times.

Second, and most importantly, for times longer than the minimum time
but smaller than a certain time, there exists an optimal control
$\kappa^{*}_{\bou}(t)$ but it is different from
$\kappa^{*}_{\unb}(t)$. This is the significant time window for ESE
protocols, since we need the connection to be possible but with the
shortest possible time. The associated time evolution for the standard
deviation comprises two branches. First, a linear branch,
$y_{1,\lin}(t)$ where $y$ denotes the position standard deviation, in
the first part of the time interval while $\kappa(t)$ has not reached
the bounds yet. Second, a branch corresponding to the solution for the
appropriate boundary value of $\kappa$ ($\kappa_{\max}$ in
compression, $0$ in decompression) in the second part of the time
interval. The two functions match smoothly, with $y_{1}(t)$ and
$\dot{y}_{1}(t)$ being continuous, at the joining time.

Rather dramatic changes are observed in the decompression case, when
the bound $\kappa\geq 0$ comes into play. This is not a mathematical
bound but a physical one: with a harmonic trap, it is experimentally
difficult to engineer a repulsive potential, and thus the stiffness
has to remain non-negative.  Most importantly, there appear two critical
values for the connection time: for $\tau_{\fin}<\tau_{d}^{(2)}$ the
bounded optimal work $\calW_{\irr,d}^{*}$ deviates from that of the
unbounded problem $\calW_{\irr,\unb}^{*}$, and at
$\tau_{\fin}=\tau_{d}^{(1)}$ we have that $\calW_{\irr,d}^{*}$ diverges.

In the last decade, stochastic heat engines have been designed by
trapping a Brownian particle in a harmonic potential, the stiffness of
which can be externally controlled
\cite{schmiedl_efficiency_2008,blickle_realization_2012,martinez_brownian_2016,ciliberto_experiments_2017},
i.e. the physical system investigated here. The cycles considered in
these Brownian heat engines typically comprise
four branches, with two of them being isothermal compression and
isothermal decompression processes. The work over these isothermal
processes must be minimised to maximise the power delivered by the
engine---the work performed by the system is minus the work performed
on the system, which is the one considered throughout this paper.

The changes in the optimal work derived here for isothermal
compression/decompression processes, which are entailed by the bounds
in the stiffness, impinge on the optimal power of the Brownian heat
engines. Specifically, the optimal power is lowered as compared with
the value obtained for unbounded stiffness.  In this respect,
analysing in detail the impact of the bounds on the power of heat
engines constitutes an interesting prospect for future research.
Another relevant venue lies in optimising mixed quantities, such as a
combination of the mean dissipated work and its standard deviation,
which may exhibit phase transitions in protocol space
\cite{Solon2018}. Also, in the realm of microfluidics
\cite{sajeesh_particle_2014}, it seems interesting to explore the
extension of the ideas presented here to the design of optimal devices
for separating and sorting particles in a desired time.

\appendix

\section{``Surgery'' method for the unbounded case}\label{app:surgery}

Here we deal with the optimisation of the work in the unbounded case
from an alternative point of view. In absence of the non-holonomic
constraint $0\leq\kappa\leq\kappa_{\max}$, one may hope to address the
optimisation problem by employing the classical variational approach
leading to the Euler-Lagrange equations. Below we show the
difficulties that arise and how to cope with them by a physically appealing ``surgery'' procedure \footnote{This ``surgery'' can
be thought of as a particular case of the procedure explained in
section 2.2 of Ref.~\cite{tolle_optimization_2012} for the Miele
problem.}.

We start by writing the irreversible work as
\begin{equation}
W_{\irr}=\int_{0}^{t_{\fin}}dt \, \dot{y}_{1}^{2},
\end{equation}
where the boundary conditions for $y_{1}$ are given by
\eqref{eq:bc-y}. Therefore, this seems to be a ``trivial'' problem:
the Euler-Lagrange equation for the optimal profile $y_{1}^{*}$ is
simply $\ddot{y}^{*}_{1}=0$
and its solution is exactly \eqref{eq:y1-linear}. The issue
arises now, because the optimal stiffness $\kappa(t)$ obtained from
\eqref{eq:y-evol-with-f1},
\begin{equation}\label{eq:kappa-appendix}
\kappa(t)=\frac{1}{y_{1}^{2}(t)}-\frac{\dot{y}_{1}(t)}{y_{1}(t)},
\end{equation}
does not verify the boundary conditions \eqref{eq:bc-k}. Note that
these boundary conditions for $\kappa$ are equivalent to
$\dot{y}_{1}(0)=\dot{y}_{1}(t_{\fin})=0$,
i.e. they ensure that the system is properly equilibrated at both the
initial and final states \footnote{The fact that the linear solution \eqref{eq:y1-linear} verifies the boundary
conditions for $y_{1}$ but not those for
$\dot{y}_{1}$ is not surprising
mathematically: the Euler-Lagrange equation is a second order
differential equation and minimises the irreversible work for given
values of $y_{1}$ at the boundaries. Then, there seems to be no room for
``extra'' boundary conditions.}.

From a physical point of view,  there should be an
optimal procedure---in the sense that the irreversible work attains a
minimum over it---to connect the initial and final equilibrium states
in a finite time. Therefore, there should be a time evolution for
$y_{1}$ that minimises the irreversible work and verifies both the
boundary conditions for $y_{1}$ and $\dot{y}_{1}$, i.e. a solution of
the \textit{overdetermined problem}
\begin{equation}\label{eq:over-determined}
\ddot{y}_{1}=0, \quad y_{1}(0)=1, \; y_{1}(t_{\fin})=y_{1,\fin}, \quad \dot{y}_{1}(0)=\dot{y}_{1}(t_{\fin})=0.
\end{equation}

Below we show that this is indeed the case by explicitly building a
solution. With this constructive procedure, what we basically reveal
is that the extra boundary conditions for $\dot{y}_{1}$ do not change
the solution in $(0,t_{\fin})$: it suffices to bring to bear the
boundary conditions for $y_{1}$ and introduce suitable jumps in
$\dot{y}_{1}$ at the boundaries. Note that we have omitted the
asterisk in the solution of the variational problem, i.e. we have
written $y_{1}$ instead of $y_{1}^{*}$ in order not to clutter
\eqref{eq:over-determined}.

To keep expressions simpler, first we introduce suitable rescalings
for both $y_{1}$ and $t$,
\begin{equation}
s\equiv \frac{t}{t_{\fin}}, \quad u\equiv \frac{y_{1}-1}{y_{1,\fin}-1},
\end{equation}
such that 
\begin{equation}
W_{\irr}=\frac{(y_{1,\fin}-1)^{2}}{t_{\fin}} \int_{0}^{1} ds \, \left(u^{\prime}\right)^{2}
\end{equation}
where the prime indicates derivative with respect to $s$, and the
overdetermined problem in \eqref{eq:over-determined} is
\begin{equation}\label{eq:over-determined-s-u}
u^{\prime\prime}=0, \quad u(0)=0, \; u(1)=1, \quad u^{\prime}(0)=u^{\prime}(1)=0.
\end{equation}

We build the family of functions in the half interval $s\in[1/2,1]$. We split the interval $[1/2,1]$ into two parts,
$[1/2,1-\epsilon]$ and $[1-\epsilon,1]$, and write down the following family of
piecewise defined functions 
\begin{subequations}
\begin{equation}
u_{\epsilon}(s)=\frac{1}{2}+\mu_{\epsilon} \left(s-\frac{1}{2}\right), \quad \frac{1}{2}\leq s\leq 1-\epsilon,
\end{equation}
\begin{align}
   u_{\epsilon}(s)=\frac{1}{2}+\mu_{\epsilon}\left(\frac{1}{2}-\epsilon\right)+
\frac{2\epsilon\mu_{\epsilon}}{\pi}\sin
  \left[\frac{\pi}{2\epsilon}\left(s-1+\epsilon\right)\right], &
    \nonumber \\
  1-\epsilon\leq s \leq 1, &
\end{align}
\end{subequations}
with
\begin{equation}
\mu_{\epsilon}=\left[1-2\epsilon\left(1-\frac{2}{\pi}\right)\right]^{-1}.
\end{equation}
It is easily shown that that the functions $u_{\epsilon}(s)$ (i)
satisfy the boundary conditions at the right endpoint $s=1$ in
\eqref{eq:over-determined-s-u} for all $\epsilon$ and (ii) are
continuous and have continuous derivative in $[1/2,1]$, including the
connection point $s=1-\epsilon$. Nevertheless, in the limit as
$\epsilon\to 0^{+}$ the ``boundary layer'' $[1-\epsilon,1]$ collapses
with $u_{\epsilon}(s)$ remaining continuous at the endpoint $s=1$ but
$u_{\epsilon}^{\prime}(s)$ becoming discontinuous. Specifically,
\begin{align}
  \lim_{s\to 1^{-}}u_{\epsilon}(s)=u(1)=1, \quad \lim_{s\to
  1^{-}}u_{\epsilon}^{\prime}(s)=1\neq u_{\epsilon}^{\prime}(1)=0,
\end{align}
because $\mu_{\epsilon}\to 1$ in the considered limit. Therefore, in
the limit as $\epsilon\to 0$ we generate the discontinuity in the
derivative of $u$---and thus of $\dot{y}_{1}$ and $\kappa$.

In the other half interval $s\in[0,1/2]$ the function is defined by a
``mirroring'' process (both left-right and up-down) with respect to
the central point $s=1/2$, $u=1/2$, i.e.
\begin{equation}
\frac{1}{2}-u_{\epsilon}(s)=u_{\epsilon}(1-s)-\frac{1}{2}, \quad 0\leq
s\leq \frac{1}{2}.
\end{equation}
The boundary conditions at $s=0$ are automatically fulfilled as a
consequence of the boundary conditions at $s=1$.

It is a matter of simple algebra to show that 
\begin{equation}
\lim_{\epsilon\to 0^{+}}\int_{0}^{1}ds \,
\left(u_{\epsilon}^{\prime}\right)^{2}=1.
\end{equation}
Therefore, the irreversible work for this family of functions
approaches the minimum value for the standard problem---with only the
values of $u$ fixed at the boundaries---as $\epsilon$ goes to
zero. Since the minimum in the overdetermined problem, with extra
conditions on the derivative, cannot be smaller than that for the
standard problem, we conclude that the solution for the overdetermined
problem is given by $\lim_{\epsilon\to 0^{+}}u_{\epsilon}(s)$. In
other words, the solution for the standard problem with a sudden
finite jump at the boundary. 

\begin{figure}
  \centering
  \includegraphics[width=3.25in]{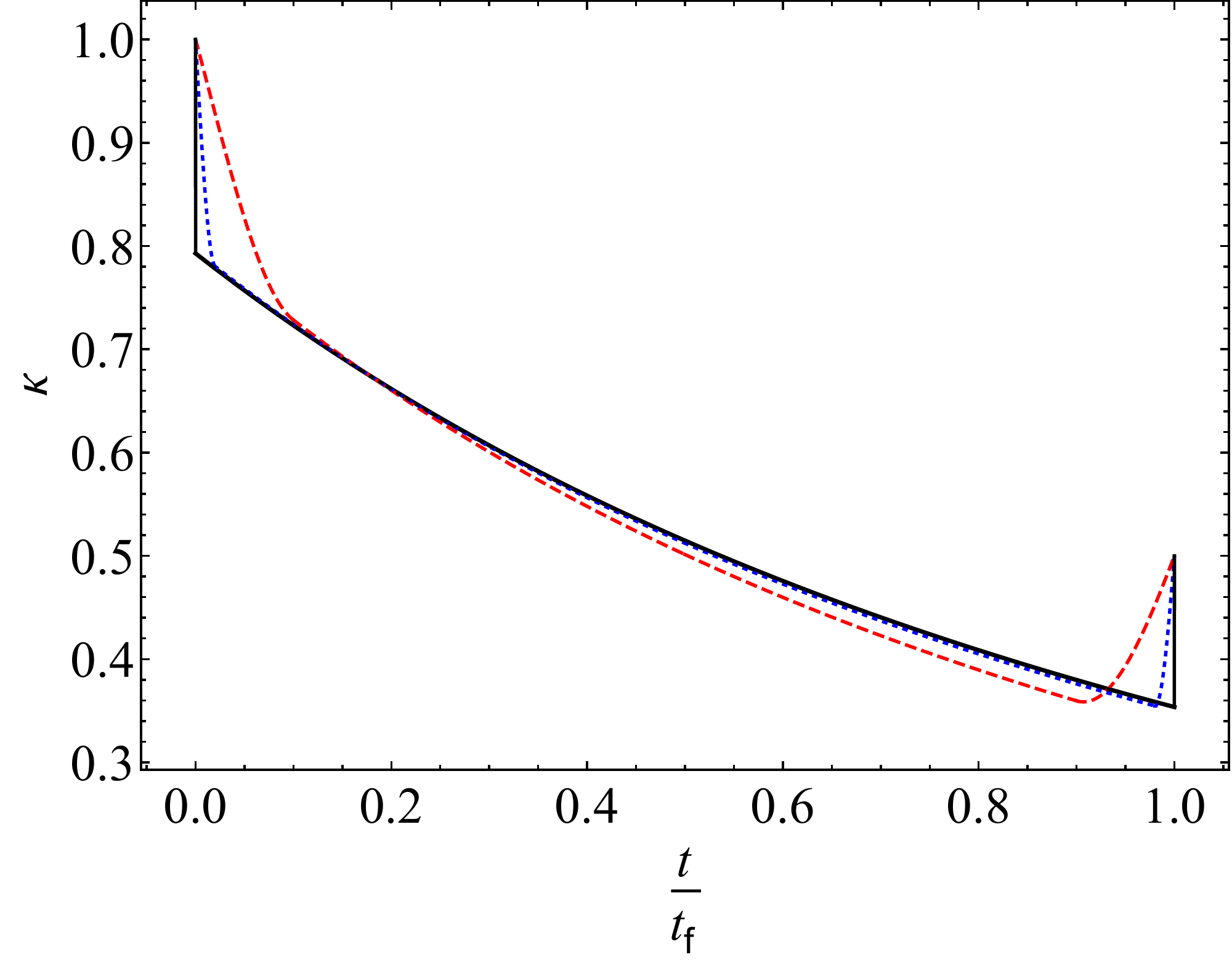}
  \caption{The surgery procedure.  Specifically, we have considered a
    decompression process with $\kappa_{\fin}=0.5$ and
    $t_{\fin}=2$. The optimal stiffness for the unbounded problem
    $\kappa_{\unb}^{*}(t)$ (solid black line), as given by
    \eqref{eq:k-unb}, is compared with the stiffness protocols
    $\kappa_{\epsilon}(t)$ stemming from \eqref{eq:kappa-appendix} for
    different values of $\epsilon$: $\epsilon=0.1$ (dashed red) and
    $\epsilon=0.02$ (dotted blue). For any value of $\epsilon$,
    $\kappa_{\epsilon}(t)$ verifies the boundary condition at
    $t=t_{\fin}$, being continuous with continuous derivative
    throughout. This is compatible with approaching
    $\kappa_{\unb}^{*}(t)$ as $\epsilon\to 0^{+}$, including the
    finite jumps at $t=0$ and $t=t_{\fin}$.}
  \label{fig:appendix}
\end{figure}

Figure~\ref{fig:appendix} shows the corresponding stiffness protocols
$\kappa_{\epsilon}(t)$, as given by inserting the family
$u_{\epsilon}(t)$ into \eqref{eq:kappa-appendix}, for several values
of $\epsilon$. They are compared with the solution
$\kappa_{\unb}^{*}(t)$ that we calculate in the main text by applying
Pontryagin's principle, which has finite jumps at the boundaries. It
is neatly observed how the proposed surgery procedure recovers the
solution $\kappa_{\unb}^{*}(t)$ in the limit as $\epsilon\to 0^{+}$,
including the jump at the boundary---although $\kappa_{\epsilon}(t)$
is continuous and has continuous derivative everywhere for any finite
$\epsilon$.

\section{Derivation of the solution for the bounded
  case}\label{app:bounded}

Let us consider the solution for the bounded case in more
detail. We focus on the decompression case because the calculations
are simpler as a consequence of our choosing $\kappa_{\min}=0$. In the
main text, we have built the optimal solution $\kappa^{*}(t)$ by
assuming that (i) when $\kappa^{*}$ touches the
boundary $\kappa=0$, then $\kappa^{*}$ remains over the boundary for
longer times, and (ii) the upper bound $\kappa_{\max}$ plays no role
in the decompression problem. In what follows, we show that this is
indeed the case.

On the one hand, the solution of the system of equations
\eqref{eq:canonical-unb} provides the optimal time evolution inside
those time windows such that the corresponding stiffness
$\tilde{\kappa}$ calculated from \eqref{eq:tilde-k} remains
non-negative, i.e.
\begin{equation}\label{eq:tildekappa-y1squared}
  \frac{\psi_{1}}{2\psi_{0}}y_{1}+1\geq 0.
\end{equation}
In those time windows, the optimal stiffness is
$\kappa^{*}=\tilde{\kappa}$ and as a consequence $y_{1}$ is a linear
function of time. Note that the left hand side of
\eqref{eq:tildekappa-y1squared} above is simply
$\tilde{\kappa}y_{1}^{2}$. On the other hand, inside the time windows
for which
\begin{equation}
\frac{\psi_{1}}{2\psi_{0}}y_{1}+1< 0,
\end{equation}
we have that $\kappa^{*}=0$ and $y_{1}(t)$ satisfies the
particularisation of \eqref{eq:y-evol-with-f1} to $\kappa=0$.

In principle, there may appear a number of different time windows with
several joining times $[0,t_{1}]$, $[t_{1},t_{2}]$, $\ldots$,
$[t_{n},t_{\fin}]$, with the solution changing from linear to the
$\kappa=0$ case (or vice versa) at each of the joining times
$t_{k}$. Now we prove that there is only one joining time
$t_{1}$ ($t_{d}^{J}$ in the notation of the main text) by
establishing that once $\tilde{\kappa}=0$ at a certain time, the
optimal control $\kappa^{*}$ remains over the boundary. To do this, we
show that
\begin{equation}
\frac{d}{dt}\left(\frac{\psi_{1}}{2\psi_{0}}y_{1}+1\right)< 0, \quad
  \text{if} \quad \frac{\psi_{1}}{2\psi_{0}}y_{1}+1 \leq 0.
\end{equation}
Since the condition $\frac{\psi_{1}}{2\psi_{0}}y_{1}+1\leq 0$ implies that
$\tilde{\kappa}\leq 0$, it suffices to prove that
$\frac{\psi_{1}}{2\psi_{0}}y_{1}+1$ is decreasing for
$\kappa=0$. By using the evolution equations for that case, it is
readily shown that
\begin{equation}
\left.\frac{d}{dt}\left(\frac{\psi_{1}}{2\psi_{0}}y_{1}+1\right)\right|_{\kappa=0}=\left[\frac{2}{y_{1}^{2}}\left(\frac{\psi_{1}}{2\psi_{0}}y_{1}+1\right)-\frac{1}{y_{1}^{2}}\right]_{\kappa=0}<0.
\end{equation}

Second, we explain why the upper bound plays no role in the
decompression process. At stated in the main text, at $t=0^{+}$ the
stiffness coming from the proposed solution is positive and lower than
$\kappa_{\ini}=1$; \eqref{eq:kappa-d-sol} leads to
$\kappa_{d}^{*}(t=0)=1-m_{d}$, and $0<m_{d}<1$. Moreover, in the
``linear'' time window $[0,t_{d}^{J}]$ the stiffness monotonically
decreases, because
\begin{equation}
\frac{d\kappa_{d}^{*}}{dt}=\frac{m_{d}}{[y_{1,\lin}(t)]^{3}}
\left[m_{d}\,y_{1,\lin}(t)-2\right]<0, \quad 0< t\leq t_{d}^{J}.
\end{equation}
Initially, this derivative is negative because $y_{1}(t=0)=1$. 
The term in parentheses increases linearly but at the
joining time $m_{d}y_{1}(t_{d}^{J})=1$, see \eqref{eq:cont-y-ydot}, so
the derivative is still negative. Therefore, it does not change sign
in the interval $[0,t_{d}^{J}]$, being always negative therein. Since
once it touches the boundary, $\kappa_{d}^{*}$ remains constant, we
have
\begin{equation}
\frac{d\kappa_{d}^{*}}{dt}\leq 0, \qquad 0< t < t_{\fin}.
\end{equation}
Thus, it is clear that the upper bound $\kappa_{\max}$ is irrelevant
when finding the optimal stiffness protocol for decompression, 
$\kappa_{d}^{*}(t)<1$ for all times.

Along similar lines, it is shown that the solution
given in the main text is the only one for compression: there is also
only one connecting time and the lower bound $\kappa_{\min}=0$ is
irrelevant in that case. The calculations are a little bit
lengthier and are thus not given here.


\begin{acknowledgments}
  C.A.P and A.P. acknowledge the support of Universidad de Sevilla’s
  VI Plan Propio de Investigación through Grant
  PP2018/494. C.A.P. also acknowledges the support from the FPU
  Fellowship Programme of the Spanish Ministerio de Educación, Cultura
  y Deporte through Grant FPU14/00241 and from STARS2018 project through UNIPD. D.G.O. and E.T. acknowledge
  financial support from the Agence Nationale de la Recherche
  (research funding Grant No. ANR-18-CE30-0013-03). E.T. also
  acknowledges funding from the Investissement d'Avenir LabEx PALM
  program (Grant No.  ANR-10-LABX-0039-PALM).
\end{acknowledgments}

\bibliography{Mi-biblioteca-22-oct-2018}

\begin{thebibliography}{39}%
\makeatletter
\providecommand \@ifxundefined [1]{%
 \@ifx{#1\undefined}
}%
\providecommand \@ifnum [1]{%
 \ifnum #1\expandafter \@firstoftwo
 \else \expandafter \@secondoftwo
 \fi
}%
\providecommand \@ifx [1]{%
 \ifx #1\expandafter \@firstoftwo
 \else \expandafter \@secondoftwo
 \fi
}%
\providecommand \natexlab [1]{#1}%
\providecommand \enquote  [1]{``#1''}%
\providecommand \bibnamefont  [1]{#1}%
\providecommand \bibfnamefont [1]{#1}%
\providecommand \citenamefont [1]{#1}%
\providecommand \href@noop [0]{\@secondoftwo}%
\providecommand \href [0]{\begingroup \@sanitize@url \@href}%
\providecommand \@href[1]{\@@startlink{#1}\@@href}%
\providecommand \@@href[1]{\endgroup#1\@@endlink}%
\providecommand \@sanitize@url [0]{\catcode `\\12\catcode `\$12\catcode
  `\&12\catcode `\#12\catcode `\^12\catcode `\_12\catcode `\%12\relax}%
\providecommand \@@startlink[1]{}%
\providecommand \@@endlink[0]{}%
\providecommand \url  [0]{\begingroup\@sanitize@url \@url }%
\providecommand \@url [1]{\endgroup\@href {#1}{\urlprefix }}%
\providecommand \urlprefix  [0]{URL }%
\providecommand \Eprint [0]{\href }%
\providecommand \doibase [0]{http://dx.doi.org/}%
\providecommand \selectlanguage [0]{\@gobble}%
\providecommand \bibinfo  [0]{\@secondoftwo}%
\providecommand \bibfield  [0]{\@secondoftwo}%
\providecommand \translation [1]{[#1]}%
\providecommand \BibitemOpen [0]{}%
\providecommand \bibitemStop [0]{}%
\providecommand \bibitemNoStop [0]{.\EOS\space}%
\providecommand \EOS [0]{\spacefactor3000\relax}%
\providecommand \BibitemShut  [1]{\csname bibitem#1\endcsname}%
\let\auto@bib@innerbib\@empty
\bibitem [{\citenamefont {Chen}\ \emph
  {et~al.}(2010{\natexlab{a}})\citenamefont {Chen}, \citenamefont {Ruschhaupt},
  \citenamefont {Schmidt}, \citenamefont {del Campo}, \citenamefont
  {Guéry-Odelin},\ and\ \citenamefont {Muga}}]{chen_fast_2010}%
  \BibitemOpen
  \bibfield  {author} {\bibinfo {author} {\bibfnamefont {X.}~\bibnamefont
  {Chen}}, \bibinfo {author} {\bibfnamefont {A.}~\bibnamefont {Ruschhaupt}},
  \bibinfo {author} {\bibfnamefont {S.}~\bibnamefont {Schmidt}}, \bibinfo
  {author} {\bibfnamefont {A.}~\bibnamefont {del Campo}}, \bibinfo {author}
  {\bibfnamefont {D.}~\bibnamefont {Guéry-Odelin}}, \ and\ \bibinfo {author}
  {\bibfnamefont {J.~G.}\ \bibnamefont {Muga}},\ }\href {\doibase
  10.1103/PhysRevLett.104.063002} {\bibfield  {journal} {\bibinfo  {journal}
  {Physical Review Letters}\ }\textbf {\bibinfo {volume} {104}},\ \bibinfo
  {pages} {063002} (\bibinfo {year} {2010}{\natexlab{a}})}\BibitemShut
  {NoStop}%
\bibitem [{\citenamefont {Chen}\ \emph
  {et~al.}(2010{\natexlab{b}})\citenamefont {Chen}, \citenamefont {Lizuain},
  \citenamefont {Ruschhaupt}, \citenamefont {Guéry-Odelin},\ and\
  \citenamefont {Muga}}]{chen_shortcut_2010}%
  \BibitemOpen
  \bibfield  {author} {\bibinfo {author} {\bibfnamefont {X.}~\bibnamefont
  {Chen}}, \bibinfo {author} {\bibfnamefont {I.}~\bibnamefont {Lizuain}},
  \bibinfo {author} {\bibfnamefont {A.}~\bibnamefont {Ruschhaupt}}, \bibinfo
  {author} {\bibfnamefont {D.}~\bibnamefont {Guéry-Odelin}}, \ and\ \bibinfo
  {author} {\bibfnamefont {J.~G.}\ \bibnamefont {Muga}},\ }\href {\doibase
  10.1103/PhysRevLett.105.123003} {\bibfield  {journal} {\bibinfo  {journal}
  {Physical Review Letters}\ }\textbf {\bibinfo {volume} {105}},\ \bibinfo
  {pages} {123003} (\bibinfo {year} {2010}{\natexlab{b}})}\BibitemShut
  {NoStop}%
\bibitem [{\citenamefont {Gu\'ery-Odelin}\ \emph {et~al.}(2014)\citenamefont
  {Gu\'ery-Odelin}, \citenamefont {Muga}, \citenamefont {Ruiz-Montero},\ and\
  \citenamefont {Trizac}}]{guery-odelin_nonequilibrium_2014}%
  \BibitemOpen
  \bibfield  {author} {\bibinfo {author} {\bibfnamefont {D.}~\bibnamefont
  {Gu\'ery-Odelin}}, \bibinfo {author} {\bibfnamefont {J.}~\bibnamefont
  {Muga}}, \bibinfo {author} {\bibfnamefont {M.}~\bibnamefont {Ruiz-Montero}},
  \ and\ \bibinfo {author} {\bibfnamefont {E.}~\bibnamefont {Trizac}},\ }\href
  {\doibase 10.1103/PhysRevLett.112.180602} {\bibfield  {journal} {\bibinfo
  {journal} {Physical Review Letters}\ }\textbf {\bibinfo {volume} {112}},\
  \bibinfo {pages} {180602} (\bibinfo {year} {2014})}\BibitemShut {NoStop}%
\bibitem [{\citenamefont {Martínez}\ \emph
  {et~al.}(2016{\natexlab{a}})\citenamefont {Martínez}, \citenamefont
  {Petrosyan}, \citenamefont {Gu\'ery-Odelin}, \citenamefont {Trizac},\ and\
  \citenamefont {Ciliberto}}]{martinez_engineered_2016}%
  \BibitemOpen
  \bibfield  {author} {\bibinfo {author} {\bibfnamefont {I.~A.}\ \bibnamefont
  {Martínez}}, \bibinfo {author} {\bibfnamefont {A.}~\bibnamefont
  {Petrosyan}}, \bibinfo {author} {\bibfnamefont {D.}~\bibnamefont
  {Gu\'ery-Odelin}}, \bibinfo {author} {\bibfnamefont {E.}~\bibnamefont
  {Trizac}}, \ and\ \bibinfo {author} {\bibfnamefont {S.}~\bibnamefont
  {Ciliberto}},\ }\href {\doibase 10.1038/nphys3758} {\bibfield  {journal}
  {\bibinfo  {journal} {Nature Physics}\ }\textbf {\bibinfo {volume} {12}},\
  \bibinfo {pages} {843} (\bibinfo {year} {2016}{\natexlab{a}})}\BibitemShut
  {NoStop}%
\bibitem [{\citenamefont {Le~Cunuder}\ \emph {et~al.}(2016)\citenamefont
  {Le~Cunuder}, \citenamefont {Martínez}, \citenamefont {Petrosyan},
  \citenamefont {Gu\'ery-Odelin}, \citenamefont {Trizac},\ and\ \citenamefont
  {Ciliberto}}]{le_cunuder_fast_2016}%
  \BibitemOpen
  \bibfield  {author} {\bibinfo {author} {\bibfnamefont {A.}~\bibnamefont
  {Le~Cunuder}}, \bibinfo {author} {\bibfnamefont {I.~A.}\ \bibnamefont
  {Martínez}}, \bibinfo {author} {\bibfnamefont {A.}~\bibnamefont
  {Petrosyan}}, \bibinfo {author} {\bibfnamefont {D.}~\bibnamefont
  {Gu\'ery-Odelin}}, \bibinfo {author} {\bibfnamefont {E.}~\bibnamefont
  {Trizac}}, \ and\ \bibinfo {author} {\bibfnamefont {S.}~\bibnamefont
  {Ciliberto}},\ }\href {\doibase 10.1063/1.4962825} {\bibfield  {journal}
  {\bibinfo  {journal} {Applied Physics Letters}\ }\textbf {\bibinfo {volume}
  {109}},\ \bibinfo {pages} {113502} (\bibinfo {year} {2016})}\BibitemShut
  {NoStop}%
\bibitem [{\citenamefont {Rondin}\ \emph {et~al.}(2017)\citenamefont {Rondin},
  \citenamefont {Gieseler}, \citenamefont {Ricci}, \citenamefont {Quidant},
  \citenamefont {Dellago},\ and\ \citenamefont
  {Novotny}}]{rondin_kramers_2017}%
  \BibitemOpen
  \bibfield  {author} {\bibinfo {author} {\bibfnamefont {L.}~\bibnamefont
  {Rondin}}, \bibinfo {author} {\bibfnamefont {J.}~\bibnamefont {Gieseler}},
  \bibinfo {author} {\bibfnamefont {F.}~\bibnamefont {Ricci}}, \bibinfo
  {author} {\bibfnamefont {R.}~\bibnamefont {Quidant}}, \bibinfo {author}
  {\bibfnamefont {C.}~\bibnamefont {Dellago}}, \ and\ \bibinfo {author}
  {\bibfnamefont {L.}~\bibnamefont {Novotny}},\ }\href {\doibase
  10.1038/nnano.2017.198} {\bibfield  {journal} {\bibinfo  {journal} {Nature
  Nanotechnology}\ }\textbf {\bibinfo {volume} {12}},\ \bibinfo {pages} {1130}
  (\bibinfo {year} {2017})}\BibitemShut {NoStop}%
\bibitem [{\citenamefont {Chupeau}\ \emph
  {et~al.}(2018{\natexlab{a}})\citenamefont {Chupeau}, \citenamefont
  {Ciliberto}, \citenamefont {Gu\'ery-Odelin},\ and\ \citenamefont
  {Trizac}}]{chupeau_engineered_2018}%
  \BibitemOpen
  \bibfield  {author} {\bibinfo {author} {\bibfnamefont {M.}~\bibnamefont
  {Chupeau}}, \bibinfo {author} {\bibfnamefont {S.}~\bibnamefont {Ciliberto}},
  \bibinfo {author} {\bibfnamefont {D.}~\bibnamefont {Gu\'ery-Odelin}}, \ and\
  \bibinfo {author} {\bibfnamefont {E.}~\bibnamefont {Trizac}},\ }\href
  {\doibase 10.1088/1367-2630/aac875} {\bibfield  {journal} {\bibinfo
  {journal} {New Journal of Physics}\ }\textbf {\bibinfo {volume} {20}},\
  \bibinfo {pages} {075003} (\bibinfo {year} {2018}{\natexlab{a}})}\BibitemShut
  {NoStop}%
\bibitem [{\citenamefont {Chupeau}\ \emph
  {et~al.}(2018{\natexlab{b}})\citenamefont {Chupeau}, \citenamefont {Besga},
  \citenamefont {Gu\'ery-Odelin}, \citenamefont {Trizac}, \citenamefont
  {Petrosyan},\ and\ \citenamefont {Ciliberto}}]{chupeau_thermal_2018}%
  \BibitemOpen
  \bibfield  {author} {\bibinfo {author} {\bibfnamefont {M.}~\bibnamefont
  {Chupeau}}, \bibinfo {author} {\bibfnamefont {B.}~\bibnamefont {Besga}},
  \bibinfo {author} {\bibfnamefont {D.}~\bibnamefont {Gu\'ery-Odelin}},
  \bibinfo {author} {\bibfnamefont {E.}~\bibnamefont {Trizac}}, \bibinfo
  {author} {\bibfnamefont {A.}~\bibnamefont {Petrosyan}}, \ and\ \bibinfo
  {author} {\bibfnamefont {S.}~\bibnamefont {Ciliberto}},\ }\href {\doibase
  10.1103/PhysRevE.98.010104} {\bibfield  {journal} {\bibinfo  {journal} {Phys.
  Rev. E}\ }\textbf {\bibinfo {volume} {98}},\ \bibinfo {pages} {010104}
  (\bibinfo {year} {2018}{\natexlab{b}})}\BibitemShut {NoStop}%
\bibitem [{\citenamefont {Sekimoto}(2010)}]{sekimoto_stochastic_2010}%
  \BibitemOpen
  \bibfield  {author} {\bibinfo {author} {\bibfnamefont {K.}~\bibnamefont
  {Sekimoto}},\ }\href@noop {} {\emph {\bibinfo {title} {Stochastic
  {Energetics}}}}\ (\bibinfo  {publisher} {Springer},\ \bibinfo {year}
  {2010})\BibitemShut {NoStop}%
\bibitem [{\citenamefont {Schmiedl}\ and\ \citenamefont
  {Seifert}(2007)}]{schmiedl_optimal_2007}%
  \BibitemOpen
  \bibfield  {author} {\bibinfo {author} {\bibfnamefont {T.}~\bibnamefont
  {Schmiedl}}\ and\ \bibinfo {author} {\bibfnamefont {U.}~\bibnamefont
  {Seifert}},\ }\href {\doibase 10.1103/PhysRevLett.98.108301} {\bibfield
  {journal} {\bibinfo  {journal} {Physical Review Letters}\ }\textbf {\bibinfo
  {volume} {98}},\ \bibinfo {pages} {108301} (\bibinfo {year}
  {2007})}\BibitemShut {NoStop}%
\bibitem [{\citenamefont {Schmiedl}\ and\ \citenamefont
  {Seifert}(2008)}]{schmiedl_efficiency_2008}%
  \BibitemOpen
  \bibfield  {author} {\bibinfo {author} {\bibfnamefont {T.}~\bibnamefont
  {Schmiedl}}\ and\ \bibinfo {author} {\bibfnamefont {U.}~\bibnamefont
  {Seifert}},\ }\href {\doibase 10.1209/0295-5075/81/20003} {\bibfield
  {journal} {\bibinfo  {journal} {EPL (Europhysics Letters)}\ }\textbf
  {\bibinfo {volume} {81}},\ \bibinfo {pages} {20003} (\bibinfo {year}
  {2008})}\BibitemShut {NoStop}%
\bibitem [{\citenamefont {Aurell}\ \emph {et~al.}(2011)\citenamefont {Aurell},
  \citenamefont {Mejía-Monasterio},\ and\ \citenamefont
  {Muratore-Ginanneschi}}]{aurell_optimal_2011}%
  \BibitemOpen
  \bibfield  {author} {\bibinfo {author} {\bibfnamefont {E.}~\bibnamefont
  {Aurell}}, \bibinfo {author} {\bibfnamefont {C.}~\bibnamefont
  {Mejía-Monasterio}}, \ and\ \bibinfo {author} {\bibfnamefont
  {P.}~\bibnamefont {Muratore-Ginanneschi}},\ }\href {\doibase
  10.1103/PhysRevLett.106.250601} {\bibfield  {journal} {\bibinfo  {journal}
  {Physical Review Letters}\ }\textbf {\bibinfo {volume} {106}},\ \bibinfo
  {pages} {250601} (\bibinfo {year} {2011})}\BibitemShut {NoStop}%
\bibitem [{\citenamefont {Band}\ \emph {et~al.}(1982)\citenamefont {Band},
  \citenamefont {Kafri},\ and\ \citenamefont {Salamon}}]{band_finite_1982}%
  \BibitemOpen
  \bibfield  {author} {\bibinfo {author} {\bibfnamefont {Y.~B.}\ \bibnamefont
  {Band}}, \bibinfo {author} {\bibfnamefont {O.}~\bibnamefont {Kafri}}, \ and\
  \bibinfo {author} {\bibfnamefont {P.}~\bibnamefont {Salamon}},\ }\href
  {\doibase 10.1063/1.329960} {\bibfield  {journal} {\bibinfo  {journal}
  {Journal of Applied Physics}\ }\textbf {\bibinfo {volume} {53}},\ \bibinfo
  {pages} {8} (\bibinfo {year} {1982})}\BibitemShut {NoStop}%
\bibitem [{\citenamefont {Tolle}(2012)}]{tolle_optimization_2012}%
  \BibitemOpen
  \bibfield  {author} {\bibinfo {author} {\bibfnamefont {H.}~\bibnamefont
  {Tolle}},\ }\href@noop {} {\emph {\bibinfo {title} {Optimization
  {Methods}}}}\ (\bibinfo  {publisher} {Springer Science \& Business Media},\
  \bibinfo {year} {2012})\BibitemShut {NoStop}%
\bibitem [{\citenamefont {Aurell}\ \emph {et~al.}(2012)\citenamefont {Aurell},
  \citenamefont {Mejía-Monasterio},\ and\ \citenamefont
  {Muratore-Ginanneschi}}]{aurell_boundary_2012}%
  \BibitemOpen
  \bibfield  {author} {\bibinfo {author} {\bibfnamefont {E.}~\bibnamefont
  {Aurell}}, \bibinfo {author} {\bibfnamefont {C.}~\bibnamefont
  {Mejía-Monasterio}}, \ and\ \bibinfo {author} {\bibfnamefont
  {P.}~\bibnamefont {Muratore-Ginanneschi}},\ }\href {\doibase
  10.1103/PhysRevE.85.020103} {\bibfield  {journal} {\bibinfo  {journal}
  {Physical Review E}\ }\textbf {\bibinfo {volume} {85}},\ \bibinfo {pages}
  {020103} (\bibinfo {year} {2012})}\BibitemShut {NoStop}%
\bibitem [{\citenamefont {Muratore-Ginanneschi}\ and\ \citenamefont
  {Schwieger}(2017)}]{muratore-ginanneschi_application_2017}%
  \BibitemOpen
  \bibfield  {author} {\bibinfo {author} {\bibfnamefont {P.}~\bibnamefont
  {Muratore-Ginanneschi}}\ and\ \bibinfo {author} {\bibfnamefont
  {K.}~\bibnamefont {Schwieger}},\ }\href {\doibase 10.3390/e19070379}
  {\bibfield  {journal} {\bibinfo  {journal} {Entropy}\ }\textbf {\bibinfo
  {volume} {19}},\ \bibinfo {pages} {379} (\bibinfo {year} {2017})}\BibitemShut
  {NoStop}%
\bibitem [{\citenamefont {Torrontegui}\ \emph {et~al.}(2013)\citenamefont
  {Torrontegui}, \citenamefont {Ibáñez}, \citenamefont {Martínez-Garaot},
  \citenamefont {Modugno}, \citenamefont {del Campo}, \citenamefont
  {Guéry-Odelin}, \citenamefont {Ruschhaupt}, \citenamefont {Chen},\ and\
  \citenamefont {Muga}}]{torrontegui_chapter_2013}%
  \BibitemOpen
  \bibfield  {author} {\bibinfo {author} {\bibfnamefont {E.}~\bibnamefont
  {Torrontegui}}, \bibinfo {author} {\bibfnamefont {S.}~\bibnamefont
  {Ibáñez}}, \bibinfo {author} {\bibfnamefont {S.}~\bibnamefont
  {Martínez-Garaot}}, \bibinfo {author} {\bibfnamefont {M.}~\bibnamefont
  {Modugno}}, \bibinfo {author} {\bibfnamefont {A.}~\bibnamefont {del Campo}},
  \bibinfo {author} {\bibfnamefont {D.}~\bibnamefont {Guéry-Odelin}}, \bibinfo
  {author} {\bibfnamefont {A.}~\bibnamefont {Ruschhaupt}}, \bibinfo {author}
  {\bibfnamefont {X.}~\bibnamefont {Chen}}, \ and\ \bibinfo {author}
  {\bibfnamefont {J.~G.}\ \bibnamefont {Muga}},\ }in\ \href {\doibase
  10.1016/B978-0-12-408090-4.00002-5} {\emph {\bibinfo {booktitle} {Advances in
  {Atomic}, {Molecular}, and {Optical} {Physics}}}},\ Vol.~\bibinfo {volume}
  {62},\ \bibinfo {editor} {edited by\ \bibinfo {editor} {\bibfnamefont
  {E.}~\bibnamefont {Arimondo}}, \bibinfo {editor} {\bibfnamefont {P.~R.}\
  \bibnamefont {Berman}}, \ and\ \bibinfo {editor} {\bibfnamefont {C.~C.}\
  \bibnamefont {Lin}}}\ (\bibinfo  {publisher} {Academic Press},\ \bibinfo
  {year} {2013})\ pp.\ \bibinfo {pages} {117--169}\BibitemShut {NoStop}%
\bibitem [{\citenamefont {Ritort}(2006)}]{ritort_single-molecule_2006}%
  \BibitemOpen
  \bibfield  {author} {\bibinfo {author} {\bibfnamefont {F.}~\bibnamefont
  {Ritort}},\ }\href
  {/citations?view_op=view_citation&continue=/scholar%3Fhl%3Den%26start%3D10%26as_sdt%3D0,5%26scilib%3D1&citilm=1&citation_for_view=fLRxLRkAAAAJ:Br1UauaknNIC&hl=en&oi=p}
  {\bibfield  {journal} {\bibinfo  {journal} {Journal of Physics: Condensed
  Matter}\ }\textbf {\bibinfo {volume} {18}},\ \bibinfo {pages} {R531}
  (\bibinfo {year} {2006})}\BibitemShut {NoStop}%
\bibitem [{\citenamefont {Wen}\ \emph {et~al.}(2007)\citenamefont {Wen},
  \citenamefont {Manosas}, \citenamefont {Li}, \citenamefont {Smith},
  \citenamefont {Bustamante}, \citenamefont {Ritort},\ and\ \citenamefont
  {Tinoco~Jr}}]{wen_force_2007}%
  \BibitemOpen
  \bibfield  {author} {\bibinfo {author} {\bibfnamefont {J.-D.}\ \bibnamefont
  {Wen}}, \bibinfo {author} {\bibfnamefont {M.}~\bibnamefont {Manosas}},
  \bibinfo {author} {\bibfnamefont {P.~T.}\ \bibnamefont {Li}}, \bibinfo
  {author} {\bibfnamefont {S.~B.}\ \bibnamefont {Smith}}, \bibinfo {author}
  {\bibfnamefont {C.}~\bibnamefont {Bustamante}}, \bibinfo {author}
  {\bibfnamefont {F.}~\bibnamefont {Ritort}}, \ and\ \bibinfo {author}
  {\bibfnamefont {I.}~\bibnamefont {Tinoco~Jr}},\ }\href
  {/citations?view_op=view_citation&continue=/scholar%3Fhl%3Den%26start%3D80%26as_sdt%3D0,5%26scilib%3D1&citilm=1&citation_for_view=fLRxLRkAAAAJ:fbc8zXXH2BUC&hl=en&oi=p}
  {\bibfield  {journal} {\bibinfo  {journal} {Biophysical Journal}\ }\textbf
  {\bibinfo {volume} {92}},\ \bibinfo {pages} {2996} (\bibinfo {year}
  {2007})}\BibitemShut {NoStop}%
\bibitem [{\citenamefont {Manosas}\ \emph {et~al.}(2007)\citenamefont
  {Manosas}, \citenamefont {Wen}, \citenamefont {Li}, \citenamefont {Smith},
  \citenamefont {Bustamante}, \citenamefont {Tinoco~Jr},\ and\ \citenamefont
  {Ritort}}]{manosas_force_2007}%
  \BibitemOpen
  \bibfield  {author} {\bibinfo {author} {\bibfnamefont {M.}~\bibnamefont
  {Manosas}}, \bibinfo {author} {\bibfnamefont {J.-D.}\ \bibnamefont {Wen}},
  \bibinfo {author} {\bibfnamefont {P.~T.~X.}\ \bibnamefont {Li}}, \bibinfo
  {author} {\bibfnamefont {S.~B.}\ \bibnamefont {Smith}}, \bibinfo {author}
  {\bibfnamefont {C.}~\bibnamefont {Bustamante}}, \bibinfo {author}
  {\bibfnamefont {I.}~\bibnamefont {Tinoco~Jr}}, \ and\ \bibinfo {author}
  {\bibfnamefont {F.}~\bibnamefont {Ritort}},\ }\href
  {/citations?view_op=view_citation&continue=/scholar%3Fhl%3Den%26start%3D80%26as_sdt%3D0,5%26scilib%3D1&citilm=1&citation_for_view=fLRxLRkAAAAJ:GFxP56DSvIMC&hl=en&oi=p}
  {\bibfield  {journal} {\bibinfo  {journal} {Biophysical Journal}\ }\textbf
  {\bibinfo {volume} {92}},\ \bibinfo {pages} {3010} (\bibinfo {year}
  {2007})}\BibitemShut {NoStop}%
\bibitem [{\citenamefont {Hoffmann}\ and\ \citenamefont
  {Dougan}(2012)}]{hoffmann_single_2012}%
  \BibitemOpen
  \bibfield  {author} {\bibinfo {author} {\bibfnamefont {T.}~\bibnamefont
  {Hoffmann}}\ and\ \bibinfo {author} {\bibfnamefont {L.}~\bibnamefont
  {Dougan}},\ }\href
  {/citations?view_op=view_citation&continue=/scholar%3Fhl%3Den%26start%3D60%26as_sdt%3D0,5%26scilib%3D1&citilm=1&citation_for_view=fLRxLRkAAAAJ:2tRrZ1ZAMYUC&hl=en&oi=p}
  {\bibfield  {journal} {\bibinfo  {journal} {Chemical Society Reviews}\
  }\textbf {\bibinfo {volume} {41}},\ \bibinfo {pages} {4781} (\bibinfo {year}
  {2012})}\BibitemShut {NoStop}%
\bibitem [{\citenamefont {Marszalek}\ and\ \citenamefont
  {Dufrêne}(2012)}]{marszalek_stretching_2012}%
  \BibitemOpen
  \bibfield  {author} {\bibinfo {author} {\bibfnamefont {P.~E.}\ \bibnamefont
  {Marszalek}}\ and\ \bibinfo {author} {\bibfnamefont {Y.~F.}\ \bibnamefont
  {Dufrêne}},\ }\href
  {/citations?view_op=view_citation&continue=/scholar%3Fhl%3Den%26start%3D70%26as_sdt%3D0,5%26scilib%3D1&citilm=1&citation_for_view=fLRxLRkAAAAJ:artPoR2Yc-kC&hl=en&oi=p}
  {\bibfield  {journal} {\bibinfo  {journal} {Chemical Society Reviews}\
  }\textbf {\bibinfo {volume} {41}},\ \bibinfo {pages} {3523} (\bibinfo {year}
  {2012})}\BibitemShut {NoStop}%
\bibitem [{\citenamefont {Ciliberto}(2017)}]{ciliberto_experiments_2017}%
  \BibitemOpen
  \bibfield  {author} {\bibinfo {author} {\bibfnamefont {S.}~\bibnamefont
  {Ciliberto}},\ }\href {\doibase 10.1103/PhysRevX.7.021051} {\bibfield
  {journal} {\bibinfo  {journal} {Physical Review X}\ }\textbf {\bibinfo
  {volume} {7}},\ \bibinfo {pages} {021051} (\bibinfo {year}
  {2017})}\BibitemShut {NoStop}%
\bibitem [{\citenamefont {Esposito}\ \emph {et~al.}(2010)\citenamefont
  {Esposito}, \citenamefont {Kawai}, \citenamefont {Lindenberg},\ and\
  \citenamefont {Van~den Broeck}}]{esposito_efficiency_2010}%
  \BibitemOpen
  \bibfield  {author} {\bibinfo {author} {\bibfnamefont {M.}~\bibnamefont
  {Esposito}}, \bibinfo {author} {\bibfnamefont {R.}~\bibnamefont {Kawai}},
  \bibinfo {author} {\bibfnamefont {K.}~\bibnamefont {Lindenberg}}, \ and\
  \bibinfo {author} {\bibfnamefont {C.}~\bibnamefont {Van~den Broeck}},\ }\href
  {\doibase 10.1103/PhysRevLett.105.150603} {\bibfield  {journal} {\bibinfo
  {journal} {Physical Review Letters}\ }\textbf {\bibinfo {volume} {105}},\
  \bibinfo {pages} {150603} (\bibinfo {year} {2010})}\BibitemShut {NoStop}%
\bibitem [{\citenamefont {Roßnagel}\ \emph {et~al.}(2014)\citenamefont
  {Roßnagel}, \citenamefont {Abah}, \citenamefont {Schmidt-Kaler},
  \citenamefont {Singer},\ and\ \citenamefont
  {Lutz}}]{rosnagel_nanoscale_2014}%
  \BibitemOpen
  \bibfield  {author} {\bibinfo {author} {\bibfnamefont {J.}~\bibnamefont
  {Roßnagel}}, \bibinfo {author} {\bibfnamefont {O.}~\bibnamefont {Abah}},
  \bibinfo {author} {\bibfnamefont {F.}~\bibnamefont {Schmidt-Kaler}}, \bibinfo
  {author} {\bibfnamefont {K.}~\bibnamefont {Singer}}, \ and\ \bibinfo {author}
  {\bibfnamefont {E.}~\bibnamefont {Lutz}},\ }\href {\doibase
  10.1103/PhysRevLett.112.030602} {\bibfield  {journal} {\bibinfo  {journal}
  {Physical Review Letters}\ }\textbf {\bibinfo {volume} {112}},\ \bibinfo
  {pages} {030602} (\bibinfo {year} {2014})}\BibitemShut {NoStop}%
\bibitem [{\citenamefont {Martínez}\ \emph
  {et~al.}(2016{\natexlab{b}})\citenamefont {Martínez}, \citenamefont
  {Roldán}, \citenamefont {Dinis}, \citenamefont {Parrondo},\ and\
  \citenamefont {Rica}}]{martinez_brownian_2016}%
  \BibitemOpen
  \bibfield  {author} {\bibinfo {author} {\bibfnamefont {I.~A.}\ \bibnamefont
  {Martínez}}, \bibinfo {author} {\bibfnamefont {E.}~\bibnamefont {Roldán}},
  \bibinfo {author} {\bibfnamefont {L.}~\bibnamefont {Dinis}}, \bibinfo
  {author} {\bibfnamefont {J.~M.~R.}\ \bibnamefont {Parrondo}}, \ and\ \bibinfo
  {author} {\bibfnamefont {R.~A.}\ \bibnamefont {Rica}},\ }\href {\doibase
  10.1038/nphys3518} {\bibfield  {journal} {\bibinfo  {journal} {Nature
  Physics}\ }\textbf {\bibinfo {volume} {12}},\ \bibinfo {pages} {67} (\bibinfo
  {year} {2016}{\natexlab{b}})}\BibitemShut {NoStop}%
\bibitem [{\citenamefont {Martínez}\ \emph {et~al.}(2017)\citenamefont
  {Martínez}, \citenamefont {Roldán}, \citenamefont {Dinis},\ and\
  \citenamefont {Rica}}]{martinez_colloidal_2017}%
  \BibitemOpen
  \bibfield  {author} {\bibinfo {author} {\bibfnamefont {I.~A.}\ \bibnamefont
  {Martínez}}, \bibinfo {author} {\bibfnamefont {E.}~\bibnamefont {Roldán}},
  \bibinfo {author} {\bibfnamefont {L.}~\bibnamefont {Dinis}}, \ and\ \bibinfo
  {author} {\bibfnamefont {R.~A.}\ \bibnamefont {Rica}},\ }\href {\doibase
  10.1039/C6SM00923A} {\bibfield  {journal} {\bibinfo  {journal} {Soft Matter}\
  }\textbf {\bibinfo {volume} {13}},\ \bibinfo {pages} {22} (\bibinfo {year}
  {2017})}\BibitemShut {NoStop}%
\bibitem [{\citenamefont {Taye}(2017)}]{taye_irreversible_2017}%
  \BibitemOpen
  \bibfield  {author} {\bibinfo {author} {\bibfnamefont {M.~A.}\ \bibnamefont
  {Taye}},\ }\href {\doibase 10.1007/s10955-017-1869-9} {\bibfield  {journal}
  {\bibinfo  {journal} {Journal of Statistical Physics}\ }\textbf {\bibinfo
  {volume} {169}},\ \bibinfo {pages} {423} (\bibinfo {year}
  {2017})}\BibitemShut {NoStop}%
\bibitem [{\citenamefont {Apertet}\ \emph {et~al.}(2017)\citenamefont
  {Apertet}, \citenamefont {Ouerdane}, \citenamefont {Goupil},\ and\
  \citenamefont {Lecoeur}}]{apertet_true_2017}%
  \BibitemOpen
  \bibfield  {author} {\bibinfo {author} {\bibfnamefont {Y.}~\bibnamefont
  {Apertet}}, \bibinfo {author} {\bibfnamefont {H.}~\bibnamefont {Ouerdane}},
  \bibinfo {author} {\bibfnamefont {C.}~\bibnamefont {Goupil}}, \ and\ \bibinfo
  {author} {\bibfnamefont {P.}~\bibnamefont {Lecoeur}},\ }\href {\doibase
  10.1103/PhysRevE.96.022119} {\bibfield  {journal} {\bibinfo  {journal}
  {Physical Review E}\ }\textbf {\bibinfo {volume} {96}},\ \bibinfo {pages}
  {022119} (\bibinfo {year} {2017})}\BibitemShut {NoStop}%
\bibitem [{\citenamefont {Blickle}\ and\ \citenamefont
  {Bechinger}(2012)}]{blickle_realization_2012}%
  \BibitemOpen
  \bibfield  {author} {\bibinfo {author} {\bibfnamefont {V.}~\bibnamefont
  {Blickle}}\ and\ \bibinfo {author} {\bibfnamefont {C.}~\bibnamefont
  {Bechinger}},\ }\href {\doibase 10.1038/nphys2163} {\bibfield  {journal}
  {\bibinfo  {journal} {Nature Physics}\ }\textbf {\bibinfo {volume} {8}},\
  \bibinfo {pages} {143} (\bibinfo {year} {2012})}\BibitemShut {NoStop}%
\bibitem [{\citenamefont {Pontryagin}(1987)}]{pontryagin_mathematical_1987}%
  \BibitemOpen
  \bibfield  {author} {\bibinfo {author} {\bibfnamefont {L.~S.}\ \bibnamefont
  {Pontryagin}},\ }\href@noop {} {\emph {\bibinfo {title} {Mathematical
  {Theory} of {Optimal} {Processes}}}}\ (\bibinfo  {publisher} {CRC Press},\
  \bibinfo {year} {1987})\BibitemShut {NoStop}%
\bibitem [{\citenamefont {Liberzon}(2012)}]{liberzon_calculus_2012}%
  \BibitemOpen
  \bibfield  {author} {\bibinfo {author} {\bibfnamefont {D.}~\bibnamefont
  {Liberzon}},\ }\href@noop {} {\emph {\bibinfo {title} {Calculus of
  {Variations} and {Optimal} {Control} {Theory}: {A} {Concise}
  {Introduction}}}}\ (\bibinfo  {publisher} {Princeton University Press},\
  \bibinfo {year} {2012})\BibitemShut {NoStop}%
\bibitem [{\citenamefont {Gelfand}\ and\ \citenamefont
  {Fomin}(2000)}]{gelfand_calculus_2000}%
  \BibitemOpen
  \bibfield  {author} {\bibinfo {author} {\bibfnamefont {I.~M.}\ \bibnamefont
  {Gelfand}}\ and\ \bibinfo {author} {\bibfnamefont {S.~V.}\ \bibnamefont
  {Fomin}},\ }\href@noop {} {\emph {\bibinfo {title} {Calculus of
  Variations}}}\ (\bibinfo  {publisher} {Dover Publications},\ \bibinfo {year}
  {2000})\BibitemShut {NoStop}%
\bibitem [{Note1()}]{Note1}%
  \BibitemOpen
  \bibinfo {note} {The main point is that the sign of all the momenta $\psi _k$
  and thus the sign of $\Pi $ can be reversed, which gives a ``mirrored''
  solution of the canonical equations. Over this ``mirrored'' solution, with
  $(-\psi _0)>0$, the corresponding $(-\Pi )$ would reach an infimum at
  $(-\protect \mathcal {H})$, instead of a supremum. It is to fix this
  ambiguity in Pontryagin\IeC {\textquoteright }s procedure and formulate a
  maximum principle that the choice $\psi _0<0$ is made~\cite
  {liberzon_calculus_2012}.}\BibitemShut {Stop}%
\bibitem [{Note2()}]{Note2}%
  \BibitemOpen
  \bibinfo {note} {This can also be understood as having a time dependent
  control set ${U}$, ${U}(0)=\kappa _{\protect \text {i}}$, ${U}(t)=[0,\kappa
  _{\protect \qopname \relax m{max}}]$ for $t\in (0,t_{\protect \text {f}})$
  and ${U}(t_{\protect \text {f}})=\kappa _{\protect \text {f}}$.}\BibitemShut
  {Stop}%
\bibitem [{\citenamefont {Solon}\ and\ \citenamefont
  {Horowitz}(2018)}]{Solon2018}%
  \BibitemOpen
  \bibfield  {author} {\bibinfo {author} {\bibfnamefont {A.~P.}\ \bibnamefont
  {Solon}}\ and\ \bibinfo {author} {\bibfnamefont {J.~M.}\ \bibnamefont
  {Horowitz}},\ }\href {\doibase 10.1103/PhysRevLett.120.180605} {\bibfield
  {journal} {\bibinfo  {journal} {Phys. Rev. Lett.}\ }\textbf {\bibinfo
  {volume} {120}},\ \bibinfo {pages} {180605} (\bibinfo {year}
  {2018})}\BibitemShut {NoStop}%
\bibitem [{\citenamefont {Sajeesh}\ and\ \citenamefont
  {Sen}(2014)}]{sajeesh_particle_2014}%
  \BibitemOpen
  \bibfield  {author} {\bibinfo {author} {\bibfnamefont {P.}~\bibnamefont
  {Sajeesh}}\ and\ \bibinfo {author} {\bibfnamefont {A.~K.}\ \bibnamefont
  {Sen}},\ }\href {\doibase 10.1007/s10404-013-1291-9} {\bibfield  {journal}
  {\bibinfo  {journal} {Microfluidics and Nanofluidics}\ }\textbf {\bibinfo
  {volume} {17}},\ \bibinfo {pages} {1} (\bibinfo {year} {2014})}\BibitemShut
  {NoStop}%
\bibitem [{Note3()}]{Note3}%
  \BibitemOpen
  \bibinfo {note} {This ``surgery'' can be thought of as a particular case of
  the procedure explained in section 2.2 of Ref.~\cite
  {tolle_optimization_2012} for the Miele problem.}\BibitemShut {Stop}%
\bibitem [{Note4()}]{Note4}%
  \BibitemOpen
  \bibinfo {note} {The fact that the linear solution \protect \textup {\hbox
  {\mathsurround \z@ \protect \normalfont (\ignorespaces \ref
  {eq:y1-linear}\unskip \@@italiccorr )}} verifies the boundary conditions for
  $y_{1}$ but not those for $\protect \mathaccentV {dot}05F{y}_{1}$ is not
  surprising mathematically: the Euler-Lagrange equation is a second order
  differential equation and minimises the irreversible work for given values of
  $y_{1}$ at the boundaries. Then, there seems to be no room for ``extra''
  boundary conditions.}\BibitemShut {Stop}%
\end{thebibliography}%

\end{document}